\documentclass{article}%
\usepackage{amsmath}
\usepackage{amsfonts}
\usepackage{amssymb}
\usepackage{graphicx}%
\newtheorem{theorem}{Theorem}

\newtheorem{corollary}[theorem]{Corollary}

\newtheorem{definition}[theorem]{Definition}

\newtheorem{proposition}[theorem]{Proposition}
\newtheorem{remark}[theorem]{Remark}

\newenvironment{proof}[1][Proof]{\noindent\textbf{#1.} }{\ \rule{0.5em}{0.5em}}
\begin{document}

\title{Clifford Valued Differential Forms, and Some Issues in Gravitation,
Electromagnetism and "Unified" Theories\thanks{Published: \textit{Int. J. Mod.
Phys. D }\textbf{13}(9), 1879-1915 (2004). This version includes corrections
of misprints and of some formulas appearing in the original text (mainly is
Section 3.5).}}
\author{W. A. Rodrigues Jr. and E. Capelas de Oliveira\\Institute of Mathematics, Statistics and Scientific Computation\\IMECC-UNICAMP CP 6065\\13083-970 Campinas-SP, Brazil\\walrod@ime.unicamp.br\hspace{0.15in}capelas@ime.unicamp.br}
\date{April 18 2004\\
updated: 16 January 2007}
\maketitle
\tableofcontents

\begin{abstract}
In this paper we show how to describe the general theory of a linear metric
compatible connection with the theory of Clifford valued differential forms.
This is done by realizing that for each spacetime point the Lie algebra of
Clifford bivectors is isomorphic to the Lie algebra of $Sl(2,\mathbb{C)}$. In
that way the pullback of the linear connection under a local trivialization of
the bundle (i.e., a choice of gauge) is represented by a Clifford valued
1-form. That observation makes it possible to realize immediately that
Einstein's gravitational theory can be formulated in a way which is similar to
a $Sl(2,\mathbb{C)}$ gauge theory. Such a theory is compared with other
interesting mathematical formulations of Einstein's theory. and particularly
with a supposedly "unified" field theory of gravitation and electromagnetism
proposed by M. Sachs. We show that his identification of Maxwell equations
within his formalism is not a valid one. Also, taking profit of the
mathematical methods introduced in the paper we investigate a very polemical
issue in Einstein gravitational theory, namely \ the problem of the
'energy-momentum' conservation. We show that many statements appearing in the
literature are confusing or even wrong.

\end{abstract}

\section{Introduction}

In this paper we introduce the concept of Clifford valued differential
forms\footnote{Analogous, but non equivalent concepts have been introduced in
\cite{dimakis,vatorr,vatorr2,tucker}. In particular \cite{dimakis} introduce
clifforms, i.e., forms with values in a abstract (internal) Clifford algebra
$\mathbb{R}_{p,q}$ associated with a pair $(\mathbb{R}^{n},g)$, where $n=p+q$
and $g$ is a bilinear form of signature $(p,q)$ in $\mathbb{R}^{n}$. These
objects \textit{differ }from the Clifford valued differential forms used in
this text., whith dispenses any abstract (internal) space.}, mathematical
entities which are sections of $\mathcal{C\ell}(TM)\otimes%
{\displaystyle\bigwedge}
T^{\ast}M$. We show how with the aid of this concept we can produce a very
beautiful \textit{description} of the theory of linear connections, where the
representative of a given linear connection in a given \textit{gauge} is given
by a \textit{bivector} valued 1-form. For that objects we introduce the
concept of \textit{exterior} covariant differential and \textit{extended}
covariant derivative operators. Our \textit{natural} definitions\footnote{Our
defintions, for the best of our knowledge, appears here for the first time.}
are to be compared with other approaches on related subjects (as described,
e.g., in \cite{baezm,beentucker,frankel,goeshu,nasen,palais,sternberg}) and
have been designed in order to parallel in a noticeable way the formalism of
the theory of connections in principal bundles and their associated covariant
derivative operators acting on associated vector bundles. We identify Cartan
curvature 2-forms and \textit{curvature bivectors}. The curvature 2-forms
satisfy Cartan's second structure equation and the curvature bivectors satisfy
equations in complete analogy with equations of gauge theories. This
immediately suggests to write Einstein's theory in that formalism, something
that has already been done and extensively studied in the past (see e.g.,
\cite{carmeli,caleini}). Our methodology however, suggests new ways of taking
advantage of such a formulation, but this is postponed for a later paper.
Here, our investigation of the $Sl(2,\mathbb{C})$ nonhomogeneous\ gauge
equation for the curvature bivector is restricted to the relationship between
that equation and some other equations which use different formalisms, but
which express the same information as the one contained in the original
Einstein's equations written in classical tensor formalism. Our analysis
includes a careful investigation of the relationship of the $Sl(2,\mathbb{C})$
nonhomogeneous\ gauge equation for the curvature bivector and some interesting
equations appearing in M. Sachs theory \cite{s1,s2,s3}.

We already showed in \cite{twospinor} that unfortunately \ M. Sachs identified
equivocally his basic variables $q_{\mu}$ as being quaternion fields over a
Lorentzian spacetime. Well, they are \textit{not}. The real mathematical
structure of these objects is that they are matrix representations of
particular sections of the even Clifford bundle of multivectors
$\mathcal{C\ell}(TM)$ (called paravector fields in mathematical literature).
Here we show that the identification proposed M. Sachs of a new antisymmetric
field \cite{s1,s2,s3} in his `unified' theory as an electromagnetic field is,
according to our opinion, an equivocated one. Indeed, as will be proved in
detail, \ M. Sachs\ `electromagnetic fields' $\mathbb{F}_{\mathbf{ab}}$ (whose
precise mathematical nature is disclosed below)\ are nothing more than some
combinations of the curvature bivectors\footnote{The curvature bivectors are
physically and mathematically equivalent to the Cartan curvature 2-forms,
since they carry the same information. This statement will become obvious from
our study in section \ 3.4.}, objects that appear naturally when we try to
formulate Einstein's gravitational theory as a $Sl(2,\mathbb{C)}$ gauge
theory.\ The equations\ found by M. Sachs, which are satisfied by his
antisymmetric fields $\mathbb{F}_{\mathbf{ab}}$ looks like Maxwell equations
in components, but they are \textit{not} Maxwell equations. However, we can
say that they do reveal one more of the many faces of Einstein's
equations\footnote{Some other faces of that equations are shown in the
Appendices.}.

Taking profit of the mathematical methods introduced in this paper we also
discuss some controversial, but conceptually important issues concerning the
law of energy-momentum conservation in General Relativity, showing that many
statements appearing in the literature are confusing and even wrong.

The paper contains two Appendices. Appendix A recalls some results of the
Clifford calculus necessary for the calculations presented in the main text.
Appendix B recalls the correct intrinsic presentation of Einstein's equations
in terms of tetrad fields $\{\theta^{\mathbf{a}}\}$, when these fields are
sections of the Clifford bundle and compare that equations, which some other
equations for that objects recently presented in the literature.

\section{Recall of Some Facts of the Theory of Linear Connections}

\subsection{Preliminaries}

In the general theory of connections \cite{choquet, konu} a connection is a
1-form in the cotangent space of a principal bundle, with values in the Lie
algebra of a gauge group. In order to develop a theory of a linear
connection\footnote{In words, \ $\overset{\blacktriangle}{%
\mbox{\boldmath{$\omega$}}%
}$ is a 1-form in the cotangent space of the bundle of ortonornal frames with
values in the Lie algebra $\mathrm{so}_{1,3}^{e}\simeq\mathrm{sl}%
(2,\mathbb{C)}$ of the group \textrm{SO}$_{_{1,3}}^{e}$. \ }
\begin{equation}
\overset{\blacktriangle}{%
\mbox{\boldmath{$\omega$}}%
}\in\sec T^{\ast}P_{\mathrm{SO}_{1,3}^{e}}(M)\otimes\mathrm{sl}(2,\mathbb{C)},
\label{9.00}%
\end{equation}
with an \textit{exterior} covariant derivative operator acting on sections of
associated vector bundles to the principal bundle $P_{\mathrm{SO}_{1,3}^{e}%
}(M)$ which reproduces moreover the well known results obtained with the usual
covariant derivative of tensor fields in the base manifold, we need to
introduce the concept of a \textit{soldering }form
\begin{equation}
\overset{\blacktriangle}{%
\mbox{\boldmath{$\theta$}}%
}\in\sec T^{\ast}P_{\mathrm{SO}_{1,3}^{e}}(M)\otimes\mathbb{R}^{1,3}.
\label{9.01}%
\end{equation}
Let be $U\subset$ $M$ and let $\varsigma:U\rightarrow\varsigma(U)\subset
P_{\mathrm{SO}_{1,3}^{e}}(M).$We are interested in the pullbacks
$\varsigma^{\ast}\overset{\blacktriangle}{%
\mbox{\boldmath{$\omega$}}%
}$ and $\varsigma^{\ast}\overset{\blacktriangle}{%
\mbox{\boldmath{$\theta$}}%
}$ once we give a local trivialization of the respective bundles. As it is
well known \cite{choquet,konu}, in a local chart $\langle x^{\mu}\rangle$
covering $U$, $\varsigma^{\ast}$ $\overset{\blacktriangle}{%
\mbox{\boldmath{$\theta$}}%
}$ uniquely determines
\begin{equation}%
\mbox{\boldmath{$\theta$}}%
=e_{\mu}\otimes dx^{\mu}\equiv e_{\mu}dx^{\mu}\in\sec TM\otimes%
{\displaystyle\bigwedge\nolimits^{1}}
T^{\ast}M. \label{9.1}%
\end{equation}

Now, we give the Clifford algebra structure to the \textit{tangent bundle},
thus generating the Clifford bundle $\mathcal{C\ell}(TM)=%
{\displaystyle\bigcup\nolimits_{x}}
\mathcal{C\ell}_{x}(M),$ with $\mathcal{C\ell}_{x}(M)\simeq\mathbb{R}_{1,3}$
introduced in Appendix A.

We recall moreover, a well known result \cite{lounesto}, namely, that for each
$x\in U\subset M$ the bivectors of $\mathcal{C\ell}(T_{x}M)$ generate under
the product defined by the commutator, the Lie algebra $\mathrm{sl}%
(2,\mathbb{C)}$. We thus are lead to define the representatives in
$\mathcal{C\ell}(TM)\otimes%
{\displaystyle\bigwedge}
T^{\ast}M$ for $%
\mbox{\boldmath{$\theta$}}%
$ and for the the pullback $%
\mbox{\boldmath{$\omega$}}%
$ of the connection \textit{in a given gauge} (that we represent with the same
symbols):
\begin{align}%
\mbox{\boldmath{$\theta$}}%
&  =e_{\mu}dx^{\mu}=\mathbf{e}_{\mathbf{a}}\theta^{\mathbf{a}}\in\sec%
{\displaystyle\bigwedge\nolimits^{1}}
TM\otimes%
{\displaystyle\bigwedge\nolimits^{1}}
T^{\ast}M\hookrightarrow\mathcal{C\ell}(TM)\otimes%
{\displaystyle\bigwedge\nolimits^{1}}
T^{\ast}M,\nonumber\\%
\mbox{\boldmath{$\omega$}}%
&  =\frac{1}{2}\omega_{\mathbf{a}}^{\mathbf{bc}}\mathbf{e}_{\mathbf{b}%
}\mathbf{e}_{\mathbf{c}}\theta^{\mathbf{a}}\nonumber\\
&  =\frac{1}{2}\omega_{\mathbf{a}}^{\mathbf{bc}}(\mathbf{e}_{\mathbf{b}}%
\wedge\mathbf{e}_{\mathbf{c}})\otimes\theta^{\mathbf{a}}\in\sec%
{\displaystyle\bigwedge\nolimits^{2}}
TM\otimes%
{\displaystyle\bigwedge\nolimits^{1}}
T^{\ast}M\hookrightarrow\mathcal{C\ell}(TM)\otimes%
{\displaystyle\bigwedge\nolimits^{1}}
T^{\ast}M. \label{9.2}%
\end{align}

Before we continue we must recall that whereas $%
\mbox{\boldmath{$\theta$}}%
$ is a true tensor, $%
\mbox{\boldmath{$\omega$}}%
$ is not a true tensor, since as it is well known, its \ `components' do not
have the tensor transformation properties. Note that the $\mathbf{\omega
}_{\mathbf{a}}^{\mathbf{bc}}$ are the \ `components' of the connection defined
by
\begin{equation}
D_{\mathbf{e}_{\mathbf{a}}}\mathbf{e}^{\mathbf{b}}=-\mathbf{\omega
}_{\mathbf{ac}}^{\mathbf{b}}\mathbf{e}^{\mathbf{c}},\hspace{0.15in}%
\mathbf{\omega}_{\mathbf{abc}}=-\mathbf{\omega}_{\mathbf{cba}}=\eta
_{\mathbf{ad}}\mathbf{\omega}_{\mathbf{bc}}^{\mathbf{d}}, \label{9.2'}%
\end{equation}
where $D_{\mathbf{e}_{\mathbf{a}}}$ is a metric compatible covariant
derivative operator\footnote{After section 3.4, $D_{\mathbf{e}_{\mathbf{a}}}$
refers to the Levi-Civita covariant derivative operator.} defined on the
tensor bundle, that naturally acts on $\mathcal{C\ell}(TM)$ (see, e.g.,
\cite{cru}). Objects like $%
\mbox{\boldmath{$\theta$}}%
$ and $%
\mbox{\boldmath{$\omega$}}%
$ will be called Clifford valued differential forms (or Clifford valued forms,
for short), and in sections 3 and 4 we give a detailed account of the algebra
and calculus of that objects. But, before we start this project we need to
recall some concepts of the theory of linear connections.

\subsection{Exterior Covariant Differential}

One of our objectives is to show how to describe, with our formalism
an\ exterior covariant differential (\textit{EXCD}) which acts naturally on
sections of Clifford valued differential forms (i.e., sections of
$\sec\mathcal{C\ell}(TM)\otimes%
{\displaystyle\bigwedge}
T^{\ast}M$ ) and which \textit{mimics} the action of the pullback of the
exterior covariant derivative operator acting on sections of a vector bundle
associated to the principal bundle $P_{\mathrm{SO}_{1,3}^{e}}(M)$, once a
linear metric compatible connection is given. Our motivation for the
definition \ of the \textit{EXCD }is that with it, the calculations of
curvature bivectors, Bianchi identities, etc., use always the same formula. Of
course, we compare our definition, with other definitions of analogous, but
distinct concepts, already used in the literature, showing where they differ
from ours, and why we think that ours seems more appropriate. In particular,
with the \textit{EXCD }and its associated \textit{extended} covariant
derivative\textit{ }(\textit{ECD}) we can write Einstein's equations in \ such
a way that the resulting equation looks like an equation for a gauge theory of
the group $Sl(2,\mathbb{C)}$. To achieve our goal, we recall below the well
known definition of the exterior covariant differential $\mathbf{d}^{E}$
acting on arbitrary sections of \ a vector bundle $E(M)$ (associated to
$P_{\mathrm{SO}_{1,3}^{e}}(M)$ and having as typical fiber a $l$-dimensional
real vector space) and on\ $\mathrm{end}E\left(  M\right)  =E\left(  M\right)
\otimes E^{\ast}(M)$, the bundle of endomorphisms of $E\left(  M\right)  $. We
recall also the concept of absolute differential acting on sections of the
tensor bundle, for the particular case of $\bigwedge\nolimits^{l}TM$.

\begin{definition}
\label{extcovop}The exterior covariant differential operator \emph{(}%
ECDO\emph{)} $\mathbf{d}^{E}$ acting on sections of $E\left(  M\right)  $ and
$\mathrm{end}E\left(  M\right)  $ is the mapping
\begin{equation}
\mathbf{d}^{E}\mathbf{:}\sec E\left(  M\right)  \rightarrow\sec E\left(
M\right)  \otimes%
{\displaystyle\bigwedge\nolimits^{1}}
T^{\ast}M,\label{W1}%
\end{equation}
such that for any differentiable function $f:M\rightarrow\mathbb{R}$,
$A\in\sec E\left(  M\right)  $ and any $F\in\sec(\mathrm{end}E\left(
M\right)  \otimes%
{\displaystyle\bigwedge\nolimits^{p}}
T^{\ast}M),$ $G\in\sec(\mathrm{end}E\left(  M\right)  \otimes%
{\displaystyle\bigwedge\nolimits^{q}}
T^{\ast}M)$ we have:%
\begin{align}
\mathbf{d}^{E}\mathbf{(}fA) &  =df\otimes A+f\mathbf{d}^{E}A,\nonumber\\
\mathbf{d}^{E}(F\otimes_{\wedge}A) &  =\mathbf{d}^{E}F\otimes_{\wedge
}A+(-1)^{p}F\otimes_{\wedge}\mathbf{d}^{E}A,\nonumber\\
\mathbf{d}^{E}(F\otimes_{\wedge}G) &  =\mathbf{d}^{E}F\otimes_{\wedge
}G+(-1)^{p}F\otimes_{\wedge}\mathbf{d}^{E}G.\label{W2}%
\end{align}

\end{definition}

In Eq.(\ref{W2}), writing $F=F^{a}\otimes f_{a}^{(p)}$, $G=G^{b}\otimes
g_{b}^{(q)}$where $F^{a}$, $G^{b}\in\sec(\mathrm{end}E\left(  M\right)  )$,
$f_{a}^{(p)}\in\sec%
{\displaystyle\bigwedge\nolimits^{p}}
T^{\ast}M$ and $g_{b}^{(q)}\in\sec%
{\displaystyle\bigwedge\nolimits^{q}}
T^{\ast}M$ we have
\begin{align}
F\otimes_{\wedge}A &  =\left(  F^{a}\otimes f_{a}^{(p)}\right)  \otimes
_{\wedge}A,\nonumber\\
F\otimes_{\wedge}G &  =\left(  F^{a}\otimes f_{a}^{(p)}\right)  \otimes
_{\wedge}G^{b}\otimes g_{b}^{(q)}.\label{W2'}%
\end{align}

In what follows, in order to simplify the notation we eventually use when
there is no possibility of confusion, the simplified (sloppy) notation%
\begin{align}
\left(  F^{a}A\right)  \otimes f_{a}^{(p)}  &  \equiv\left(  F^{a}A\right)
f_{a}^{(p)},\nonumber\\
\left(  F^{a}\otimes f_{a}^{(p)}\right)  \otimes_{\wedge}G^{b}\otimes
g_{b}^{(q)}  &  =\left(  F^{a}G^{b}\right)  f_{a}^{(p)}\wedge g_{b}^{\left(
q\right)  }, \label{W2''}%
\end{align}
\ where $F^{a}A\in\sec E\left(  M\right)  $ and $F^{a}G^{b}$means the
composition of the respective endomorphisms.

Let $U\subset M$ be an open subset of $M$, $\langle x^{\mu}\rangle$ a
coordinate functions of a maximal atlas of $M$, $\{e_{\mu}\}$ a coordinate
basis of $TU\subset TM$ and $\{s_{\mathbf{K}}\},$ $\mathbf{K=}1,2,...l$ a
basis for any $\sec E\left(  U\right)  \subset\sec E\left(  M\right)  $. Then,
a basis for any section of $E\left(  M\right)  \otimes%
{\displaystyle\bigwedge\nolimits^{1}}
T^{\ast}M$ is given by $\{s_{\mathbf{K}}\otimes dx^{\mu}\}$.

\begin{definition}
The covariant derivative operator $D_{e_{\mu}}:\sec E\left(  M\right)
\rightarrow\sec E\left(  M\right)  $ is given by
\begin{equation}
\mathbf{d}^{E}A\doteq\left(  D_{e_{\mu}}A\right)  \otimes dx^{\mu}, \label{W3}%
\end{equation}
where, writing $A=A^{\mathbf{K}}\otimes s_{\mathbf{K}}$ we have%
\begin{equation}
D_{e_{\mu}}A=\partial_{\mu}A^{\mathbf{K}}\otimes s_{\mathbf{K}}+A^{\mathbf{K}%
}\otimes D_{e_{\mu}}s_{\mathbf{K}}. \label{W4}%
\end{equation}

\end{definition}

Now, let examine the case where $E\left(  M\right)  =TM\equiv%
{\displaystyle\bigwedge\nolimits^{1}}
(TM)\hookrightarrow\mathcal{C\ell}(TM)$. Let $\{\mathbf{e}_{\mathbf{j}}\},$ be
an orthonormal basis of $TM$. Then,using Eq.(\ref{W4}) and Eq.(\ref{9.2'})
\begin{align}
\mathbf{d}^{E}\mathbf{e}_{\mathbf{j}} &  =(D_{e_{\mathbf{k}}}\mathbf{e}%
_{\mathbf{j}})\otimes\theta^{\mathbf{k}}\equiv\mathbf{e}_{\mathbf{k}}\otimes%
\mbox{\boldmath{$\omega$}}%
_{\mathbf{j}}^{\mathbf{k}}\nonumber\\%
\mbox{\boldmath{$\omega$}}%
_{\mathbf{j}}^{\mathbf{k}} &  =%
\mbox{\boldmath{$\omega$}}%
_{\mathbf{rj}}^{\mathbf{k}}\theta^{\mathbf{r}},\label{W5}%
\end{align}
where the $%
\mbox{\boldmath{$\omega$}}%
_{\mathbf{j}}^{\mathbf{k}}\in\sec%
{\displaystyle\bigwedge\nolimits^{1}}
T^{\ast}M$ are the so-called \textit{connection 1-forms}.

Also, for $\mathbf{v=}v^{\mathbf{i}}\mathbf{e}_{\mathbf{i}}\in\sec TM$, we
have
\begin{align}
\mathbf{d}^{E}\mathbf{v}  &  =D_{\mathbf{e}_{\mathbf{i}}}\mathbf{v\otimes
}\theta^{\mathbf{i}}=\mathbf{e}_{\mathbf{i}}\otimes\mathbf{d}^{E}%
v^{\mathbf{i}},\nonumber\\
\mathbf{d}^{E}v^{\mathbf{i}}  &  =dv^{\mathbf{i}}+%
\mbox{\boldmath{$\omega$}}%
_{\mathbf{k}}^{\mathbf{i}}v^{\mathbf{k}}. \label{9.5}%
\end{align}

\subsection{Absolute Differential}

\ Now, \ let \ $E\left(  M\right)  =TM\equiv%
{\displaystyle\bigwedge\nolimits^{l}}
(TM)\hookrightarrow\mathcal{C\ell}(TM).$\ Recall that the usual
\textit{absolute differential} $D$ of $A\in\sec%
{\displaystyle\bigwedge\nolimits^{l}}
TM\hookrightarrow\sec\mathcal{C\ell}\left(  TM\right)  $ is a mapping (see,
e.g., \cite{choquet})%
\begin{equation}
D\mathbf{:}\sec%
{\displaystyle\bigwedge\nolimits^{l}}
TM\rightarrow\sec%
{\displaystyle\bigwedge\nolimits^{l}}
TM\otimes%
{\displaystyle\bigwedge\nolimits^{1}}
T^{\ast}M, \label{W.12}%
\end{equation}
such that for any differentiable $A\in\sec%
{\displaystyle\bigwedge\nolimits^{l}}
TM$ we have%
\begin{equation}
DA=\left(  D_{\mathbf{e}_{\mathbf{i}}}A\right)  \otimes\theta^{i},
\label{W.14}%
\end{equation}
where $D_{e_{\mathbf{i}}}A$ is the standard covariant derivative of $A\in\sec%
{\displaystyle\bigwedge\nolimits^{l}}
TM\hookrightarrow\sec\mathcal{C\ell}\left(  TM\right)  $. Also, for any
differentiable function $f:M\rightarrow\mathbb{R}$, and differentiable
$A\in\sec%
{\displaystyle\bigwedge\nolimits^{l}}
TM$ we have%

\begin{equation}
D\mathbf{(}fA)=df\otimes A+fDA. \label{9.13}%
\end{equation}

Now, if we suppose that the orthonormal basis $\{\mathbf{e}_{\mathbf{j}}\}$ of
$TM$ is such that each $\mathbf{e}_{\mathbf{j}}\in\sec%
{\displaystyle\bigwedge\nolimits^{1}}
TM$ $\hookrightarrow\sec\mathcal{C\ell}\left(  TM\right)  ,$ we can find
easily using the Clifford algebra structure of the space of multivectors that
Eq.(\ref{W5}) can be written as:
\begin{align}
D\mathbf{e}_{\mathbf{j}}  &  =(D_{\mathbf{e}_{\mathbf{k}}}\mathbf{e}%
_{\mathbf{j}})\theta^{\mathbf{k}}=\frac{1}{2}[%
\mbox{\boldmath{$\omega$}}%
,\mathbf{e}_{\mathbf{j}}]=-\mathbf{e}_{\mathbf{j}}\lrcorner%
\mbox{\boldmath{$\omega$}}%
\nonumber\\%
\mbox{\boldmath{$\omega$}}%
&  =\frac{1}{2}\omega_{\mathbf{k}}^{\mathbf{ab}}\mathbf{e}_{\mathbf{a}}%
\wedge\mathbf{e}_{\mathbf{b}}\otimes\theta^{\mathbf{k}}\nonumber\\
&  \equiv\frac{1}{2}\omega_{\mathbf{k}}^{\mathbf{ab}}\mathbf{e}_{\mathbf{a}%
}\mathbf{e}_{\mathbf{b}}\otimes\theta^{\mathbf{k}}\in\sec%
{\displaystyle\bigwedge\nolimits^{2}}
TM\otimes%
{\displaystyle\bigwedge\nolimits^{1}}
T^{\ast}M\hookrightarrow\sec\mathcal{C\ell}(TM)\otimes%
{\displaystyle\bigwedge\nolimits^{1}}
T^{\ast}M, \label{9.9}%
\end{align}
where $%
\mbox{\boldmath{$\omega$}}%
$ is the \textit{representative} of the connection in a given gauge.

The general case is given by the following proposition.

\begin{proposition}
For $A\in\sec%
{\displaystyle\bigwedge\nolimits^{l}}
TM\hookrightarrow\sec\mathcal{C\ell}\left(  TM\right)  $ we have%
\begin{equation}
DA=dA+\frac{1}{2}[%
\mbox{\boldmath{$\omega$}}%
,A]. \label{W.15}%
\end{equation}

\end{proposition}

\begin{proof}
The proof is a simple calculation, left to the reader.
\end{proof}

Eq.(\ref{W.15}) can now be extended by linearity for an arbitrary
nonhomogeneous multivector $A\in\sec\mathcal{C\ell}\left(  TM\right)
$.\medskip

\begin{remark}
We see that when $E(M)=%
{\displaystyle\bigwedge\nolimits^{l}}
TM\hookrightarrow\sec\mathcal{C\ell}\left(  TM\right)  $ the absolute
differential $D$ can be identified with the exterior covariant derivative
$\mathbf{d}^{E}$.
\end{remark}

We proceed now to find an appropriate\textit{ exterior }covariant differential
which acts naturally on Clifford valued differential forms, i.e., objects that
are sections of \ $\mathcal{C\ell}(TM)\otimes%
{\displaystyle\bigwedge}
T^{\ast}M$ $(\equiv%
{\displaystyle\bigwedge}
T^{\ast}M\ \otimes\mathcal{C\ell}(TM))$ (see next section).\ Note that we
cannot simply use the above definition by using $E\left(  M\right)
=\mathcal{C\ell}(TM)$ and \textrm{end}$E\left(  M\right)  =$ \textrm{end}%
$\mathcal{C\ell}(TM)$, because \textrm{end}$\mathcal{C\ell}(TM)\neq
\mathcal{C\ell}(TM)\otimes%
{\displaystyle\bigwedge}
T^{\ast}M$. Instead, we must use the above theory and possible applications as
a guide in order to find an appropriate definition. Let us see how this can be done.

\section{Clifford Valued Differential Forms}

\begin{definition}
A \textit{homogeneous} multivector valued differential form of type $(l,p)$ is
a section of $%
{\displaystyle\bigwedge\nolimits^{l}}
TM\otimes%
{\displaystyle\bigwedge\nolimits^{p}}
T^{\ast}M\hookrightarrow\mathcal{C\ell}(TM)\otimes%
{\displaystyle\bigwedge}
T^{\ast}M$, for $0\leq l\leq4$, $0\leq p\leq4$. A section of $\mathcal{C\ell
}(TM)\otimes%
{\displaystyle\bigwedge}
T^{\ast}M$ such that the multivector part is non homogeneous is called a
Clifford valued differential form.
\end{definition}

We recall, that any $A\in\sec%
{\displaystyle\bigwedge\nolimits^{l}}
TM\otimes%
{\displaystyle\bigwedge\nolimits^{p}}
T^{\ast}M$ $\hookrightarrow\sec\mathcal{C\ell}(TM)\otimes%
{\displaystyle\bigwedge\nolimits^{p}}
T^{\ast}M$ can always be written as%
\begin{align}
A  &  =m_{(l)}\otimes\psi^{(p)}\equiv\frac{1}{l!}m_{(l)}^{\mathbf{i}%
_{1}...\mathbf{i}_{l}}\mathbf{e}_{\mathbf{i}_{1}}...\mathbf{e}_{\mathbf{i}%
_{l}}\otimes\psi^{(p)}\nonumber\\
&  =\frac{1}{p!}m_{(l)}\otimes\psi_{\mathbf{j}_{1}...\mathbf{j}_{p}}%
^{(p)}\theta^{\mathbf{j}_{1}}\wedge...\wedge\theta^{\mathbf{j}_{p}}\nonumber\\
&  =\frac{1}{l!p!}m_{(l)}^{\mathbf{i}_{1}...\mathbf{i}_{l}}\mathbf{e}%
_{\mathbf{i}_{1}}...\mathbf{e}_{\mathbf{i}_{l}}\otimes\psi_{\mathbf{j}%
_{1}....\mathbf{j}_{p}}^{(p)}\theta^{\mathbf{j}_{1}}\wedge...\wedge
\theta^{\mathbf{j}_{p}}\label{9.6}\\
&  =\frac{1}{l!p!}A_{\mathbf{j}_{1}...\mathbf{j}_{p}}^{\mathbf{i}%
_{1}...\mathbf{i}_{l}}\mathbf{e}_{\mathbf{i}_{1}}...\mathbf{e}_{\mathbf{i}%
_{l}}\otimes\theta^{\mathbf{i}_{1}}\wedge...\wedge\theta^{\mathbf{i}_{p}%
}.\nonumber
\end{align}
\medskip

\begin{definition}
The $\otimes_{\wedge}$ product of $A=$ $\overset{m}{A}\otimes\psi^{(p)}\in
\sec\mathcal{C\ell}(TM)\otimes%
{\displaystyle\bigwedge\nolimits^{p}}
T^{\ast}M$ and $B=$ $\overset{m}{B}\otimes\chi^{(p)}\in\sec\mathcal{C\ell
}(TM)\otimes%
{\displaystyle\bigwedge\nolimits^{q}}
T^{\ast}M$ is the mapping\footnote{$\overset{m}{A}$ and $\overset{m}{B}$ are
general nonhomogeous multivector fields.}: \
\begin{align}
\otimes_{\wedge} &  :\sec\mathcal{C\ell}(TM)\otimes%
{\displaystyle\bigwedge\nolimits^{l}}
T^{\ast}M\times\sec\mathcal{C\ell}(TM)\otimes%
{\displaystyle\bigwedge\nolimits^{p}}
T^{\ast}M\nonumber\\
&  \rightarrow\sec\mathcal{C\ell}(TM)\otimes%
{\displaystyle\bigwedge\nolimits^{l+p}}
T^{\ast}M,\nonumber\\
A\otimes_{\wedge}B &  =\text{ }\overset{m}{A}\overset{m}{B}\otimes
\psi^{\left(  p\right)  }\wedge\chi^{(q)}.\label{9.6PROD}%
\end{align}
\smallskip
\end{definition}

\begin{definition}
The \textit{commutator} $[A,B]$ of \ $A\in\sec%
{\displaystyle\bigwedge\nolimits^{l}}
TM\otimes%
{\displaystyle\bigwedge\nolimits^{p}}
T^{\ast}M\hookrightarrow\sec\mathcal{C\ell}(TM)\otimes%
{\displaystyle\bigwedge\nolimits^{p}}
T^{\ast}M$ and $B\in%
{\displaystyle\bigwedge\nolimits^{m}}
TM\otimes%
{\displaystyle\bigwedge\nolimits^{q}}
T^{\ast}M\hookrightarrow\sec\mathcal{C\ell}(TM)\otimes%
{\displaystyle\bigwedge\nolimits^{q}}
T^{\ast}M$\ \ is the mapping:%
\begin{align}
\lbrack\hspace{0.15in},\hspace{0.15in}]  &  :\sec%
{\displaystyle\bigwedge\nolimits^{l}}
TM\otimes%
{\displaystyle\bigwedge\nolimits^{p}}
T^{\ast}M\times\sec%
{\displaystyle\bigwedge\nolimits^{m}}
TM\otimes%
{\displaystyle\bigwedge\nolimits^{q}}
T^{\ast}M\nonumber\\
&  \rightarrow\sec((%
{\displaystyle\sum\limits_{k=|l-m|}^{|l+m|}}
{\displaystyle\bigwedge\nolimits^{k}}
T^{\ast}M)\otimes%
{\displaystyle\bigwedge\nolimits^{p+q}}
T^{\ast}M)\nonumber\\
\lbrack A,B]  &  =A\otimes_{\wedge}B-\left(  -1\right)  ^{pq}B\otimes_{\wedge
}A \label{9.7}%
\end{align}
Writing $A=\frac{1}{l!}A^{\mathbf{j}_{1}..\mathbf{.j}_{l}}\mathbf{e}%
_{\mathbf{j}_{1}}\mathbf{...e}_{\mathbf{j}_{l}}\psi^{\left(  p\right)  }$,
$B=\frac{1}{m!}B^{\mathbf{i}_{1}...\mathbf{i}_{m}}\mathbf{e}_{\mathbf{i}_{1}%
}\mathbf{...e}_{\mathbf{i}_{m}}\chi^{(q)}$, with $\psi^{(p)}\in\sec%
{\displaystyle\bigwedge\nolimits^{p}}
T^{\ast}M$ and $\chi^{(q)}\in\sec%
{\displaystyle\bigwedge\nolimits^{q}}
T^{\ast}M$, we have
\begin{equation}
\lbrack A,B]=\frac{1}{l!m!}A^{\mathbf{j}_{1}..\mathbf{.j}_{l}}B^{\mathbf{i}%
_{1}...\mathbf{i}_{m}}\left[  \mathbf{e}_{\mathbf{j}_{1}}\mathbf{...e}%
_{\mathbf{j}_{l}}\mathbf{,e}_{\mathbf{i}_{1}}\mathbf{...e}_{\mathbf{i}_{m}%
}\right]  \psi^{\left(  p\right)  }\wedge\chi^{(q)}. \label{9.7BIS}%
\end{equation}
\ The definition of\ the commutator is extended by linearity to arbitrary
sections of $\mathcal{C\ell}(TM)\otimes%
{\displaystyle\bigwedge}
T^{\ast}M$.
\end{definition}

Now, we have the proposition.

\begin{proposition}
Let \ $A\in\sec\mathcal{C\ell}(TM)\otimes%
{\displaystyle\bigwedge\nolimits^{p}}
T^{\ast}M$, $B\in\sec\mathcal{C\ell}(TM)\otimes%
{\displaystyle\bigwedge\nolimits^{q}}
T^{\ast}M$, $C\in A\in\sec\mathcal{C\ell}(TM)\otimes%
{\displaystyle\bigwedge\nolimits^{r}}
T^{\ast}M$. Then,%
\begin{equation}
\lbrack A,B]=(-1)^{1+pq}[B,A], \label{p1}%
\end{equation}
and%
\begin{equation}
(-1)^{pr}\left[  \left[  A,B\right]  ,C\right]  +(-1)^{qp}\left[  \left[
B,C\right]  ,A\right]  +(-)^{rq}\left[  \left[  C,A\right]  ,B\right]  =0.
\label{p2}%
\end{equation}

\end{proposition}

\begin{proof}
It follows directly from a simple calculation, left to the reader.
\end{proof}

Eq.(\ref{p2}) may be called the \textit{graded Jacobi identity }%
\cite{bleecker}.

\begin{corollary}
Let be $A^{(2)}\in\sec%
{\displaystyle\bigwedge\nolimits^{2}}
(TM)\otimes%
{\displaystyle\bigwedge\nolimits^{p}}
T^{\ast}M$ and \ $B\in\sec%
{\displaystyle\bigwedge\nolimits^{r}}
(TM)\otimes%
{\displaystyle\bigwedge\nolimits^{q}}
T^{\ast}M$. Then,%
\begin{equation}
\lbrack A^{(2)},B]=C, \label{p2bis}%
\end{equation}
where $C\in\sec%
{\displaystyle\bigwedge\nolimits^{r}}
(TM)\otimes%
{\displaystyle\bigwedge\nolimits^{p+q}}
T^{\ast}M$.
\end{corollary}

\begin{proof}
It follows from \ a direct \ calculation, left to the reader.$\blacksquare$

\begin{proposition}
Let \ $\omega\in\sec%
{\displaystyle\bigwedge\nolimits^{2}}
(TM)\otimes%
{\displaystyle\bigwedge\nolimits^{1}}
T^{\ast}M$ , $A\in\sec%
{\displaystyle\bigwedge\nolimits^{l}}
(TM)\otimes%
{\displaystyle\bigwedge\nolimits^{p}}
T^{\ast}M$.$B\in\sec%
{\displaystyle\bigwedge\nolimits^{m}}
(TM)\otimes%
{\displaystyle\bigwedge\nolimits^{q}}
T^{\ast}M$. Then, we have%
\begin{equation}
\lbrack\omega,A\otimes_{\wedge}B]=[\omega,A]\otimes_{\wedge}B+(-1)^{p}%
A\otimes_{\wedge}[\omega,B]. \label{p.2eureka}%
\end{equation}
\ 
\end{proposition}

\hspace{-0.4cm}\textbf{Proof}. Using the definition of the commutator we can
write%
\begin{align*}
\lbrack%
\mbox{\boldmath{$\omega$}}%
,A]\otimes_{\wedge}B  &  =(%
\mbox{\boldmath{$\omega$}}%
\otimes_{\wedge}A-(-1)^{p}A\otimes_{\wedge}%
\mbox{\boldmath{$\omega$}}%
)\otimes_{\wedge}B\\
&  =(%
\mbox{\boldmath{$\omega$}}%
\otimes_{\wedge}A\otimes_{\wedge}B-(-1)^{p+q}A\otimes_{\wedge}B\otimes
_{\wedge}%
\mbox{\boldmath{$\omega$}}%
)\\
&  +(-1)^{p+q}A\otimes_{\wedge}B\otimes_{\wedge}%
\mbox{\boldmath{$\omega$}}%
-(-1)^{p}A\otimes_{\wedge}%
\mbox{\boldmath{$\omega$}}%
\otimes_{\wedge}B\\
&  =[%
\mbox{\boldmath{$\omega$}}%
,A\otimes_{\wedge}B]-(-1)^{p}A\otimes_{\wedge}[%
\mbox{\boldmath{$\omega$}}%
,B],
\end{align*}

from where the desired result follows.
\end{proof}

From Eq.(\ref{13.p.2eureka}) we have also\footnote{The result printed in the
original printed version is (unfortunately) wrong. However (fortunately)
except for details, it does not change any of the conclusions.}
\begin{align*}
&  (p+q)[%
\mbox{\boldmath{$\omega$}}%
,A\otimes_{\wedge}B]\\
&  =p[%
\mbox{\boldmath{$\omega$}}%
,A]\otimes_{\wedge}B+(-1)^{p}qA\otimes_{\wedge}[%
\mbox{\boldmath{$\omega$}}%
,B]\\
&  +q[%
\mbox{\boldmath{$\omega$}}%
,A]\otimes_{\wedge}B+(-1)^{p}pA\otimes_{\wedge}[%
\mbox{\boldmath{$\omega$}}%
,B].
\end{align*}

\begin{definition}
The action of the differential operator $d$ acting on%
\[
A\in\sec%
{\displaystyle\bigwedge\nolimits^{l}}
TM\otimes%
{\displaystyle\bigwedge\nolimits^{p}}
T^{\ast}M\hookrightarrow\sec\mathcal{C\ell}(TM)\otimes%
{\displaystyle\bigwedge\nolimits^{p}}
T^{\ast}M,
\]
is given by:%
\begin{align}
dA  &  \circeq\mathbf{e}_{\mathbf{j}_{1}}...\mathbf{e}_{\mathbf{j}_{l}}\otimes
dA^{\mathbf{j}_{1}...\mathbf{j}_{l}}\label{9.8}\\
&  =\mathbf{e}_{\mathbf{j}_{1}}...\mathbf{e}_{\mathbf{j}_{l}}\otimes d\frac
{1}{p!}A_{\mathbf{i}_{1}...\mathbf{i}_{p}}^{\mathbf{j}_{1}...\mathbf{j}_{l}%
}\theta^{\mathbf{i}_{1}}\wedge...\wedge\theta^{\mathbf{i}_{p}}.\nonumber
\end{align}
\ 
\end{definition}

We have the important\medskip\ proposition.

\begin{proposition}
Let $A\in\sec\mathcal{C\ell}(TM)\otimes%
{\displaystyle\bigwedge\nolimits^{p}}
T^{\ast}M$ and $B\in\sec\mathcal{C\ell}(TM)\otimes%
{\displaystyle\bigwedge\nolimits^{q}}
T^{\ast}M$. Then,
\begin{equation}
d[A,B]=[dA,B]+(-1)^{p}[A,dB]. \label{p3}%
\end{equation}

\end{proposition}

\begin{proof}
The proof is a simple calculation, left to the reader.
\end{proof}

We now \ define the exterior covariant differential operator (\textit{EXCD)}
$\mathbf{D}$ and the \textit{extended} covariant derivative (\textit{ECD})
$\mathbf{D}_{\mathbf{e}_{\mathbf{r}}}$ acting on a Clifford valued form
$\ \mathcal{A}\in\sec%
{\displaystyle\bigwedge\nolimits^{l}}
TM\otimes%
{\displaystyle\bigwedge\nolimits^{p}}
T^{\ast}M\hookrightarrow\sec\mathcal{C\ell}\left(  TM\right)  $ $\otimes%
{\displaystyle\bigwedge\nolimits^{p}}
T^{\ast}M$, as follows.

\subsection{Exterior Covariant Differential of Clifford Valued Forms}

\begin{definition}
The exterior covariant differential of $\mathcal{A}$ is the mapping :%
\begin{align}
\mathbf{D}  &  \mathbf{:}\sec%
{\displaystyle\bigwedge\nolimits^{l}}
TM\otimes%
{\displaystyle\bigwedge\nolimits^{p}}
T^{\ast}M\rightarrow\sec[(%
{\displaystyle\bigwedge\nolimits^{l}}
TM\otimes%
{\displaystyle\bigwedge\nolimits^{p}}
T^{\ast}M)\otimes_{\wedge}%
{\displaystyle\bigwedge\nolimits^{1}}
T^{\ast}M]\nonumber\\
&  \subset\sec%
{\displaystyle\bigwedge\nolimits^{l}}
TM\otimes%
{\displaystyle\bigwedge\nolimits^{p+1}}
T^{\ast}M,\nonumber\\
\mathbf{D}\mathcal{A}  &  =d\mathcal{A}+\frac{p}{2}[%
\mbox{\boldmath{$\omega$}}%
,\mathcal{A}],\text{ if \ }\mathcal{A\in}\sec%
{\displaystyle\bigwedge\nolimits^{l}}
TM\otimes%
{\displaystyle\bigwedge\nolimits^{p}}
T^{\ast}M,\text{ }l,p\geq1. \label{W.21BIS}%
\end{align}

\end{definition}

\begin{proposition}
Let $\mathcal{A}\in\sec%
{\displaystyle\bigwedge\nolimits^{l}}
TM\otimes%
{\displaystyle\bigwedge\nolimits^{p}}
T^{\ast}M\hookrightarrow\sec\mathcal{C\ell}\left(  TM\right)  $ $\otimes%
{\displaystyle\bigwedge\nolimits^{p}}
T^{\ast}M$, $\mathcal{B}\in\sec%
{\displaystyle\bigwedge\nolimits^{m}}
TM\otimes%
{\displaystyle\bigwedge\nolimits^{q}}
T^{\ast}M\hookrightarrow\sec\mathcal{C\ell}\left(  TM\right)  $ $\otimes%
{\displaystyle\bigwedge\nolimits^{q}}
T^{\ast}M$. Then, the exterior differential satisfies
\begin{align}
\mathbf{D}(\mathcal{A}\otimes_{\wedge}\mathcal{B)}  &  \mathcal{=}%
\mathbf{D}\mathcal{A}\otimes_{\wedge}\mathcal{B+}(-1)^{p}\mathcal{A}%
\otimes_{\wedge}\mathbf{D}\mathcal{B}\nonumber\\
&  +q[%
\mbox{\boldmath{$\omega$}}%
,\mathcal{A}]\otimes_{\wedge}\mathcal{B+}(-1)^{p}p\mathcal{A}\otimes_{\wedge}[%
\mbox{\boldmath{$\omega$}}%
,\mathcal{B}]. \label{W.16biss}%
\end{align}

\end{proposition}

\begin{proof}
It follows directly from the definition if we take into account the properties
of the product \ $\otimes_{\wedge}$ and Eq.(\ref{p.2eureka}).\footnote{Observe
that $\mathbf{D}$ does not satisfy the Leibniz rule, contrary to what was
stated in the original printed version. However, fortunately this does not
change any conclusion.}
\end{proof}

\subsection{Extended Covariant Derivative of Clifford Valued Forms}

\begin{definition}
\ The \textit{extended covariant derivative operator is the mapping}%
\[
\mathbf{D}_{e_{\mathbf{r}}}:\sec%
{\displaystyle\bigwedge\nolimits^{l}}
TM\otimes%
{\displaystyle\bigwedge\nolimits^{p}}
T^{\ast}M\rightarrow\sec%
{\displaystyle\bigwedge\nolimits^{l}}
TM\otimes%
{\displaystyle\bigwedge\nolimits^{p}}
T^{\ast}M,
\]
\textbf{ }such that for any $\mathcal{A}\in\sec%
{\displaystyle\bigwedge\nolimits^{l}}
TM\otimes%
{\displaystyle\bigwedge\nolimits^{p}}
T^{\ast}M\hookrightarrow\sec\mathcal{C\ell}\left(  TM\right)  $ $\otimes%
{\displaystyle\bigwedge\nolimits^{p}}
T^{\ast}M$, $l,$ $p\geq1$, we have
\end{definition}

\begin{equation}
\mathbf{D}\mathcal{A=(}\mathbf{D}_{\mathbf{e}_{\mathbf{r}}}\mathcal{A)\otimes
}_{\wedge}\theta^{r}. \label{EXTCODER}%
\end{equation}

We can immediately verify that
\begin{equation}
\mathbf{D}_{\mathbf{e}_{\mathbf{r}}}\mathcal{A=}\mathbf{\partial
}_{e_{\mathbf{r}}}\mathcal{A}+\frac{p}{2}[%
\mbox{\boldmath{$\omega$}}%
_{\mathbf{r}},\mathcal{A}],
\end{equation}
and, of course, in general\footnote{For a Clifford algebra formula for the
calculation of $D_{e_{\mathbf{r}}}\mathcal{A}$, $\mathcal{A\in}\sec%
{\displaystyle\bigwedge\nolimits^{p}}
T^{\ast}M$ see Eq.(\ref{der1}).
\par
{}} \
\begin{equation}
\mathbf{D}_{\mathbf{e}_{\mathbf{r}}}\mathcal{A\neq}D_{e_{\mathbf{r}}%
}\mathcal{A}. \label{OK}%
\end{equation}

Let us write explicitly some important cases which will appear latter.

\subsubsection{Case\textbf{ }$p=1$}

Let $\mathcal{A}\in\sec%
{\displaystyle\bigwedge\nolimits^{l}}
TM\otimes%
{\displaystyle\bigwedge\nolimits^{1}}
T^{\ast}M\hookrightarrow\sec\mathcal{C\ell}\left(  TM\right)  $ $\otimes%
{\displaystyle\bigwedge\nolimits^{1}}
T^{\ast}M$. Then,
\begin{equation}
\mathbf{D}\mathcal{A}=d\mathcal{A}+\frac{1}{2}[%
\mbox{\boldmath{$\omega$}}%
,\mathcal{A}], \label{18}%
\end{equation}
or%
\begin{equation}
\mathbf{D}_{e_{\mathbf{k}}}\mathcal{A}=\partial_{e_{\mathbf{r}}}%
\mathcal{A}+\frac{1}{2}[%
\mbox{\boldmath{$\omega$}}%
_{\mathbf{k}},\mathcal{A}]. \label{W.18a}%
\end{equation}

\subsubsection{Case $p=2$}

\ Let $\mathcal{F}\in\sec%
{\displaystyle\bigwedge\nolimits^{l}}
TM\otimes%
{\displaystyle\bigwedge\nolimits^{2}}
T^{\ast}M\hookrightarrow\sec\mathcal{C\ell}\left(  TM\right)  $ $\otimes%
{\displaystyle\bigwedge^{2}}
T^{\ast}M$. Then, \
\begin{equation}
\mathbf{D}\mathcal{F}=d\mathcal{F}+[%
\mbox{\boldmath{$\omega$}}%
,\mathcal{F}], \label{W.19}%
\end{equation}

\begin{equation}
\mathbf{D}_{e_{\mathbf{r}}}\mathcal{F}=\partial_{e_{\mathbf{r}}}\mathcal{F}+[%
\mbox{\boldmath{$\omega$}}%
_{\mathbf{r}},\mathcal{F}]. \label{W.19bis}%
\end{equation}

\subsection{Cartan Exterior Differential}

Recall that \cite{frankel} \textit{Cartan} defined the exterior covariant
differential $\ $of $\mathfrak{C}=e_{\mathbf{i}}\otimes\mathfrak{C}%
^{\mathbf{i}}\in$\ sec$%
{\displaystyle\bigwedge\nolimits^{1}}
TM\otimes%
{\displaystyle\bigwedge\nolimits^{p}}
T^{\ast}M$ as a mapping%
\begin{align}
\mathbf{D}^{c}  &  \mathbf{:}\
{\displaystyle\bigwedge\nolimits^{1}}
TM\otimes%
{\displaystyle\bigwedge\nolimits^{p}}
T^{\ast}M\longrightarrow\
{\displaystyle\bigwedge\nolimits^{1}}
TM\otimes%
{\displaystyle\bigwedge\nolimits^{p+1}}
T^{\ast}M,\nonumber\\
\mathbf{D}^{c}\mathfrak{C}  &  =\mathbf{D}^{c}\mathbf{(}e_{\mathbf{i}}%
\otimes\mathfrak{C}^{\mathbf{i}})=e_{\mathbf{i}}\otimes d\mathfrak{C}%
^{\mathbf{i}}+\mathbf{D}^{c}e_{\mathbf{i}}\wedge\mathfrak{C}^{\mathbf{i}%
},\label{cartan1}\\
\mathbf{D}^{c}e_{\mathbf{j}}  &  =(D_{e_{\mathbf{k}}}e_{\mathbf{j}}%
)\theta^{\mathbf{k}}\nonumber
\end{align}
which in view of Eq.(\ref{9.8}) and Eq.(\ref{9.9}) can be written as
\begin{equation}
\mathbf{D}^{c}\mathfrak{C}=\mathbf{D}^{c}\mathbf{(}e_{\mathbf{i}}%
\otimes\mathfrak{C}^{\mathbf{i}})=d\mathfrak{C}+\frac{1}{2}[%
\mbox{\boldmath{$\omega$}}%
,\mathfrak{C}]. \label{cartan2}%
\end{equation}

So, we have, for $p>1$, the following relation between the exterior covariant
differential $\mathbf{D}$ and Cartan's exterior differential ($p>1$)
\begin{equation}
\mathbf{D}\mathfrak{C=}\mathbf{D}^{c}\mathfrak{C}+\frac{p-1}{2}[%
\mbox{\boldmath{$\omega$}}%
,\mathfrak{C}]. \label{cartan3}%
\end{equation}

Note moreover that when $\mathfrak{C}^{(1)}=e_{\mathbf{i}}\otimes
\mathfrak{C}^{\mathbf{i}}\in$\ sec$%
{\displaystyle\bigwedge\nolimits^{1}}
TM\otimes%
{\displaystyle\bigwedge\nolimits^{1}}
T^{\ast}M$, we have
\begin{equation}
\mathbf{D}\mathfrak{C}^{(1)}\mathfrak{=}\mathbf{D}^{c}\mathfrak{C}^{(1)}.
\label{cartan4}%
\end{equation}

We end this section with two observations:

(i) There are other approaches to the concept of exterior covariant
differential acting on sections of a vector bundle $E\otimes%
{\displaystyle\bigwedge\nolimits^{p}}
T^{\ast}M$ and also in sections of $\mathrm{end}(E)$ $\otimes%
{\displaystyle\bigwedge\nolimits^{p}}
T^{\ast}M$, as e.g., in
\cite{baezm,beentucker,frankel,goeshu,nasen,palais,sternberg}. Not all are
completely equivalent among themselves and to the one presented above. Our
definitions, we think, have the merit of mimicking coherently the pullback
under a local section of the covariant differential acting on sections of
vector bundles associated to a given principal bundle as used in gauge
theories. Indeed, this consistence will be checked in several situations below.

(ii) Some authors, e.g., \cite{beentucker,thirring2} find convenient to
introduce the concept of \textit{exterior covariant derivative} of
\textit{indexed} $p$-forms, which are objects like the curvature $2$-forms
(see below) \textit{but not} the connection 1-forms introduced above. We do
not use such concept in this paper.

\subsection{Torsion and Curvature}

Let $%
\mbox{\boldmath{$\theta$}}%
=e_{\mu}dx^{\mu}=\mathbf{e}_{\mathbf{a}}\theta^{\mathbf{a}}\in\sec%
{\displaystyle\bigwedge\nolimits^{1}}
TM\otimes%
{\displaystyle\bigwedge\nolimits^{1}}
T^{\ast}M\hookrightarrow\mathcal{C\ell}(TM)\otimes%
{\displaystyle\bigwedge\nolimits^{1}}
T^{\ast}M$ \ and \ $%
\mbox{\boldmath{$\omega$}}%
=\frac{1}{2}\left(  \omega_{\mathbf{a}}^{\mathbf{bc}}\mathbf{e}_{\mathbf{b}%
}\wedge\mathbf{e}_{\mathbf{c}}\right)  \otimes\theta^{\mathbf{a}}\equiv
\frac{1}{2}\omega_{\mathbf{a}}^{\mathbf{bc}}\mathbf{e}_{\mathbf{b}}%
\mathbf{e}_{\mathbf{c}}\theta^{\mathbf{a}}\in\sec%
{\displaystyle\bigwedge\nolimits^{2}}
M\otimes%
{\displaystyle\bigwedge\nolimits^{1}}
T^{\ast}M\hookrightarrow\mathcal{C\ell}(TM)\otimes%
{\displaystyle\bigwedge\nolimits^{1}}
T^{\ast}M$ \ be respectively the \textit{representatives} of a soldering form
and a connection on the \textit{basis manifold. }Then, following the standard
procedure \cite{konu}, the \textit{torsion} of the connection and the
\textit{curvature} of the connection on the basis manifold are defined by
\begin{equation}%
\mbox{\boldmath{$\Theta$}}%
=\mathbf{D}%
\mbox{\boldmath{$\theta$}}%
\in\sec%
{\displaystyle\bigwedge\nolimits^{1}}
TM\otimes%
{\displaystyle\bigwedge\nolimits^{2}}
T^{\ast}M\hookrightarrow\mathcal{C\ell}(TM)\otimes%
{\displaystyle\bigwedge\nolimits^{2}}
T^{\ast}M, \label{9.15}%
\end{equation}
and
\begin{equation}
\mathcal{R=}\mathbf{D}%
\mbox{\boldmath{$\omega$}}%
\in\sec%
{\displaystyle\bigwedge\nolimits^{2}}
M\otimes%
{\displaystyle\bigwedge\nolimits^{2}}
T^{\ast}M\hookrightarrow\mathcal{C\ell}(TM)\otimes%
{\displaystyle\bigwedge\nolimits^{2}}
T^{\ast}M. \label{9.16}%
\end{equation}
We now calculate $%
\mbox{\boldmath{$\Theta$}}%
$ and $\mathbf{D}\mathcal{R}$. We have,%

\begin{equation}
\mathbf{D}%
\mbox{\boldmath{$\theta$}}%
=\mathbf{D(}\mathbf{e}_{\mathbf{a}}\theta^{\mathbf{a}})=\mathbf{e}%
_{\mathbf{a}}d\theta^{\mathbf{a}}+\frac{1}{2}[%
\mbox{\boldmath{$\omega$}}%
_{\mathbf{a}},\mathbf{e}_{\mathbf{d}}]\theta^{\mathbf{a}}\wedge\theta
^{\mathbf{d}} \label{9.17}%
\end{equation}
and since $\frac{1}{2}[%
\mbox{\boldmath{$\omega$}}%
_{\mathbf{a}},\mathbf{e}_{\mathbf{d}}]=-\mathbf{e}_{\mathbf{d}}\lrcorner%
\mbox{\boldmath{$\omega$}}%
_{\mathbf{a}}=\omega_{\mathbf{ad}}^{\mathbf{c}}\mathbf{e}_{\mathbf{c}}$ we have%

\begin{equation}
\mathbf{D(}\mathbf{e}_{\mathbf{a}}\theta^{\mathbf{a}})=\mathbf{e}_{\mathbf{a}%
}[d\theta^{\mathbf{a}}+\omega_{\mathbf{bd}}^{\mathbf{a}}\theta^{\mathbf{b}%
}\wedge\theta^{\mathbf{d}}]=\mathbf{e}_{\mathbf{a}}%
\mbox{\boldmath{$\Theta$}}%
^{\mathbf{a}}, \label{9.18}%
\end{equation}
and we recognize%
\begin{equation}%
\mbox{\boldmath{$\Theta$}}%
^{\mathbf{a}}=d\theta^{\mathbf{a}}+\omega_{\mathbf{bd}}^{\mathbf{a}}%
\theta^{\mathbf{b}}\wedge\theta^{\mathbf{d}}, \label{9.19}%
\end{equation}
as \textit{Cartan's first structure equation}.

For a torsion free connection, the torsion 2-forms $%
\mbox{\boldmath{$\Theta$}}%
^{\mathbf{a}}=0$, and it follows that $%
\mbox{\boldmath{$\Theta$}}%
=0$. A metric compatible connection $%
\mbox{\boldmath{$\omega$}}%
$ (for which $D_{\mathbf{e}_{\mathbf{a}}}g=0,\mathbf{a}=0,1,2,3$) satisfying $%
\mbox{\boldmath{$\Theta$}}%
^{\mathbf{a}}=0$ is called a Levi-Civita connection. In the remaining of this
paper we \textit{restrict} ourself to that case.

Now, according to Eq.(\ref{W.21BIS}) we have,
\begin{equation}
\mathbf{D}\mathcal{R}=d\mathcal{R}+[%
\mbox{\boldmath{$\omega$}}%
,\mathcal{R}]. \label{9.20}%
\end{equation}

Now, taking into account that
\begin{equation}
\mathcal{R}=d%
\mbox{\boldmath{$\omega$}}%
+\frac{1}{2}[%
\mbox{\boldmath{$\omega$}}%
,%
\mbox{\boldmath{$\omega$}}%
], \label{9.21}%
\end{equation}
and that from Eqs.(\ref{p1}).(\ref{p2}) and (\ref{p3}) it follows that%

\begin{align}
d[%
\mbox{\boldmath{$\omega$}}%
,%
\mbox{\boldmath{$\omega$}}%
]  &  =[d%
\mbox{\boldmath{$\omega$}}%
,%
\mbox{\boldmath{$\omega$}}%
]-[%
\mbox{\boldmath{$\omega$}}%
,d%
\mbox{\boldmath{$\omega$}}%
],\nonumber\\
\lbrack d%
\mbox{\boldmath{$\omega$}}%
,%
\mbox{\boldmath{$\omega$}}%
]  &  =-[%
\mbox{\boldmath{$\omega$}}%
,d%
\mbox{\boldmath{$\omega$}}%
],\nonumber\\
\lbrack\lbrack%
\mbox{\boldmath{$\omega$}}%
,%
\mbox{\boldmath{$\omega$}}%
],%
\mbox{\boldmath{$\omega$}}%
]  &  =0, \label{9.22}%
\end{align}
we have immediately%
\begin{equation}
\mathbf{D}\mathcal{R}=d\mathcal{R}+[%
\mbox{\boldmath{$\omega$}}%
,\mathcal{R}]=0. \label{9.23}%
\end{equation}

Eq.(\ref{9.23}) is known as the \textit{Bianchi identity}.

Note that, since $\{\mathbf{e}_{\mathbf{a}}\}$ is an orthonormal frame we can
write:
\begin{align}
\mathcal{R}  &  =\frac{1}{4}R_{\mu\nu}^{\mathbf{ab}}\mathbf{e}_{\mathbf{a}%
}\wedge\mathbf{e}_{\mathbf{b}}\otimes(dx^{\mu}\wedge dx^{\nu})\nonumber\\
&  \equiv\frac{1}{4}\mathcal{R}_{\mathbf{cd}}^{\mathbf{ab}}\mathbf{e}%
_{\mathbf{a}}\mathbf{e}_{\mathbf{b}}\otimes\theta^{\mathbf{c}}\wedge
\theta^{\mathbf{d}}=\frac{1}{4}R_{\rho\sigma}^{\alpha\beta}e_{\mathbf{\alpha}%
}e_{\beta}\otimes dx^{\rho}\wedge dx^{\mathbf{\sigma}}\nonumber\\
&  =\frac{1}{4}R_{\mathcal{\mu\nu\rho\sigma}}e^{\mu}e^{\nu}\otimes dx^{\rho
}\wedge dx^{\mathbf{\sigma}}, \label{9.24}%
\end{align}
where $R_{\mathcal{\mu\nu\rho\sigma}}$ are the components of the curvature
tensor, also known in differential geometry as the Riemann tensor. We recall
the well known symmetries%
\begin{align}
R_{\mathcal{\mu\nu\rho\sigma}}  &  =-R_{\mathcal{\nu\mu\rho\sigma}%
},\nonumber\\
R_{\mathcal{\mu\nu\rho\sigma}}  &  =-R_{\mathcal{\mu\nu\sigma\rho}%
},\nonumber\\
R_{\mathcal{\mu\nu\rho\sigma}}  &  =R_{\mathcal{\rho\sigma\mu\nu}}.
\label{9.42}%
\end{align}

We also write Eq.(\ref{9.24}) as
\begin{align}
\mathcal{R} &  =\frac{1}{4}R_{\mathbf{cd}}^{\mathbf{ab}}\mathbf{e}%
_{\mathbf{a}}\mathbf{e}_{\mathbf{b}}\otimes(\theta^{\mathbf{c}}\wedge
\theta^{\mathbf{d}})=\frac{1}{2}\mathbf{R}_{\mu\nu}dx^{\mu}\wedge
dx^{\mathbf{\nu}}\nonumber\\
&  =\frac{1}{2}\mathcal{R}_{\mathbf{b}\ }^{\mathbf{a}}\mathbf{e}_{\mathbf{a}%
}\mathbf{e}^{\mathbf{b}},\label{9.43}%
\end{align}
with
\begin{align}
\mathbf{R}_{\mu\nu} &  =\frac{1}{2}R_{\mu\nu}^{\mathbf{ab}}\mathbf{e}%
_{\mathbf{a}}\mathbf{e}_{\mathbf{b}}=\frac{1}{2}R_{\mu\nu}^{\mathbf{ab}%
}\mathbf{e}_{\mathbf{a}}\wedge\mathbf{e}_{\mathbf{b}}\in\sec%
{\displaystyle\bigwedge\nolimits^{2}}
TM\hookrightarrow\mathcal{C\ell}(TM),\nonumber\\
\mathcal{R}^{\mathbf{ab}} &  =\frac{1}{2}R_{\mu\nu}^{\mathbf{ab}}dx^{\mu
}\wedge dx^{\mathbf{\nu}}\in\sec%
{\displaystyle\bigwedge\nolimits^{2}}
T^{\ast}M,\label{9.44}%
\end{align}
where $\mathbf{R}_{\mu\nu}$ will be called curvature bivectors and the
$\mathcal{R}_{\mathbf{b}}^{\mathbf{a}}$ are called after Cartan the curvature
$2$-forms. The $\mathcal{R}_{\mathbf{b}}^{\mathbf{a}}$ satisfy
\textit{Cartan's second structure equation}%
\begin{equation}
\mathcal{R}_{\mathbf{b}}^{\mathbf{a}}=d%
\mbox{\boldmath{$\omega$}}%
_{\mathbf{b}}^{\mathbf{a}}+%
\mbox{\boldmath{$\omega$}}%
_{\mathbf{c}}^{\mathbf{a}}\wedge%
\mbox{\boldmath{$\omega$}}%
_{\mathbf{d}}^{\mathbf{c}},\label{9.45}%
\end{equation}
which follows calculating $d\mathcal{R}$ from Eq.(\ref{9.21}). \ Now, we can
also write,
\begin{align}
\mathbf{D}\mathcal{R} &  =d\mathcal{R}+[%
\mbox{\boldmath{$\omega$}}%
,\mathcal{R}]\nonumber\\
&  =\frac{1}{2}\{d(\frac{1}{2}R_{\mu\nu}^{\mathbf{ab}}\mathbf{e}_{\mathbf{a}%
}\mathbf{e}_{\mathbf{b}}dx^{\mu}\wedge dx^{\nu})+[%
\mbox{\boldmath{$\omega$}}%
_{\rho},\mathbf{R}_{\mu\nu}]\}dx^{\rho}\wedge dx^{\mu}\wedge dx^{\nu
}\nonumber\\
&  =\frac{1}{2}\{\partial_{\rho}\mathbf{R}_{\mu\nu}+[%
\mbox{\boldmath{$\omega$}}%
_{\rho},\mathbf{R}_{\mu\nu}]\}dx^{\rho}\wedge dx^{\mu}\wedge dx^{\nu
}\label{9.25}\\
&  =\frac{1}{2}\mathbf{D}_{e_{\rho}}\mathbf{R}_{\mu\nu}dx^{\rho}\wedge
dx^{\mu}\wedge dx^{\nu}\nonumber\\
&  =\frac{1}{3!}\left(  \mathbf{D}_{e_{\rho}}\mathbf{R}_{\mu\nu}%
+\mathbf{D}_{e_{\mu}}\mathbf{R}_{\nu\rho}+\mathbf{D}_{e_{\nu}}\mathbf{R}%
_{\rho\mu}\right)  dx^{\rho}\wedge dx^{\mu}\wedge dx^{\nu}=0,\nonumber
\end{align}
\textit{ }\ from where it follows that
\begin{equation}
\mathbf{D}_{e_{\rho}}\mathbf{R}_{\mu\nu}+\mathbf{D}_{e_{\mu}}\mathbf{R}%
_{\nu\rho}+\mathbf{D}_{e_{\nu}}\mathbf{R}_{\rho\mu}=0.\label{9.28}%
\end{equation}

\begin{remark}
Eq.(\ref{9.28}) \ is called in Physics textbooks on gauge theories (see, e.g.,
\cite{nasen,ryder}) Bianchi identity. Note that physicists call the extended
covariant derivative operator
\begin{equation}
\mathbf{D}_{e_{\rho}}\equiv\mathbf{D}_{\rho}=\partial_{\rho}+[%
\mbox{\boldmath{$\omega$}}%
_{\rho},], \label{9.26}%
\end{equation}
acting on the curvature bivectors as the \ `\textit{covariant derivative'.
Note however that, as detailed above, this operator is not the usual covariant
derivative operator }$D_{\mathbf{e}_{a}}$ acting on sections of the tensor bundle.
\end{remark}

We now find the explicit expression for the curvature bivectors $\mathbf{R}%
_{\mu\nu}$ in terms \ of the connections bivectors $%
\mbox{\boldmath{$\omega$}}%
_{\mu}=%
\mbox{\boldmath{$\omega$}}%
(e_{\mu})$,\ which will be used latter. First recall that by definition%
\begin{equation}
\mathbf{R}_{\mu\nu}=\mathcal{R}(e_{\mu},e_{\nu})=-\mathcal{R}(e_{\nu},e_{\mu
})=-\mathbf{R}_{\mu\nu}. \label{9.29}%
\end{equation}

Now, observe that using Eqs.(\ref{p1}), (\ref{p2}) and (\ref{p3}) we can
easily show that%
\begin{align}
\lbrack%
\mbox{\boldmath{$\omega$}}%
,%
\mbox{\boldmath{$\omega$}}%
](e_{\mu},e_{\nu})  &  =2[%
\mbox{\boldmath{$\omega$}}%
(e_{\mu}),%
\mbox{\boldmath{$\omega$}}%
(e_{\nu})]\nonumber\\
&  =2[%
\mbox{\boldmath{$\omega$}}%
_{\mu},%
\mbox{\boldmath{$\omega$}}%
_{\nu}]. \label{9.30}%
\end{align}

Using Eqs. (\ref{9.21}), (\ref{9.29}) and (\ref{9.30}) we get
\begin{equation}
\mathbf{R}_{\mu\nu}=\partial_{\mu}%
\mbox{\boldmath{$\omega$}}%
_{v}-\partial_{v}%
\mbox{\boldmath{$\omega$}}%
_{\mu}+[%
\mbox{\boldmath{$\omega$}}%
_{\mu},%
\mbox{\boldmath{$\omega$}}%
_{\nu}]. \label{9.40}%
\end{equation}

\subsection{Some Useful Formulas}

Let $A\in\sec%
{\displaystyle\bigwedge\nolimits^{p}}
TM\hookrightarrow\sec\mathcal{C\ell}(TM)$ and $\mathcal{R}$ the curvature of
the connection \ as defined in Eq.(\ref{9.16}). Then\footnote{In the original
printed version it is unfortunately written $\mathbf{D}^{2}A=\frac{1}%
{2}[\mathcal{R},A]$.} as a detailed calculation can show,
\begin{equation}
\mathbf{D}^{2}A=\frac{1}{4}[\mathcal{R},A]+\frac{1}{4}[d%
\mbox{\boldmath{$\omega$}}%
,A]. \label{9.T1}%
\end{equation}

Also, we can show using the previous result that if \ $\mathcal{A}\in
\sec\mathcal{C\ell}(TM)\otimes%
{\displaystyle\bigwedge\nolimits^{1}}
T^{\ast}M$ it holds\footnote{In the printed version it is unfortunately
written that $\mathbf{D}^{2}\mathcal{A}=\frac{1}{2}[\mathcal{R},\mathcal{A}%
]$.}%
\begin{equation}
\mathbf{D}^{2}\mathcal{A}=\frac{1}{4}[\mathcal{R},\mathcal{A}]+\frac{1}{2}[%
\mbox{\boldmath{$\omega$}}%
,dA]. \label{9.TC1}%
\end{equation}

\section{General Relativity as a $Sl\left(  2,\mathbb{C}\right)  $ Gauge
Theory}

\subsection{The Nonhomogeneous Field Equations}

The analogy of the fields $\mathbf{R}_{\mu\nu}=\frac{1}{2}R_{\mu\nu
}^{\mathbf{ab}}\mathbf{e}_{\mathbf{a}}\mathbf{e}_{\mathbf{b}}=\frac{1}%
{2}R_{\mu\nu}^{\mathbf{ab}}\mathbf{e}_{\mathbf{a}}\wedge\mathbf{e}%
_{\mathbf{b}}\in\sec%
{\displaystyle\bigwedge\nolimits^{2}}
TM\hookrightarrow\mathcal{C\ell}(TM)$ with the gauge fields of particle fields
is so appealing that it is irresistible to propose some kind of a
$Sl(2,\mathbb{C)}$ formulation for the gravitational field. And indeed this
has already been done, and the interested reader may consult, e.g.,
\cite{carmeli,mielke}. \ Here, we observe that despite the similarities, the
gauge theories of particle physics are in general formulated in flat Minkowski
spacetime and the theory here must be for a field on a general Lorentzian
spacetime. This introduces additional complications, but it is not our purpose
to discuss that issue with all attention it deserves here. Indeed, for our
purposes in this paper we will need only to recall some facts.

To start, recall that in gauge theories besides the homogenous field equations
given by Bianchi's identities, we also have the nonhomogeneous field equation.
This equation, in analogy to the nonhomogeneous equation for the
electromagnetic field (see Eq.(\ref{1.9}) in Appendix A) is written here as%

\begin{equation}
\mathbf{D\star}\mathcal{R}=d\mathbf{\star}\mathcal{R+}[%
\mbox{\boldmath{$\omega$}}%
,\mathbf{\star}\mathcal{R}]=-\star\mathcal{J}\mathbf{,}\label{10.0}%
\end{equation}
where the $\mathcal{J}\in\sec%
{\displaystyle\bigwedge\nolimits^{2}}
TM\otimes%
{\displaystyle\bigwedge\nolimits^{1}}
T^{\ast}M\hookrightarrow\mathcal{C\ell}(TM)\otimes%
{\displaystyle\bigwedge\nolimits^{1}}
T^{\ast}M$ is a `current', which, if the theory is to be one equivalent to
General Relativity, must be in some way related with the energy momentum
tensor in Einstein theory. In order to write from this equation an equation
for the curvature bivectors, it is very useful to imagine that \ $%
{\displaystyle\bigwedge}
T^{\ast}M\hookrightarrow\mathcal{C\ell}(T^{\ast}M)$, the Clifford bundle of
differential forms, for in that case the powerful calculus described in the
Appendix A can be used. So, we write:
\begin{align}%
\mbox{\boldmath{$\omega$}}%
&  \in\sec%
{\displaystyle\bigwedge\nolimits^{2}}
TM\otimes%
{\displaystyle\bigwedge\nolimits^{1}}
T^{\ast}M\hookrightarrow\mathcal{C\ell}(TM)\otimes%
{\displaystyle\bigwedge\nolimits^{1}}
T^{\ast}M\hookrightarrow\mathcal{C\ell}(TM)\otimes\mathcal{C\ell}(T^{\ast
}M),\nonumber\\
\mathcal{R} &  =\mathbf{D}%
\mbox{\boldmath{$\omega$}}%
\in\sec%
{\displaystyle\bigwedge\nolimits^{2}}
TM\otimes%
{\displaystyle\bigwedge\nolimits^{2}}
T^{\ast}M\hookrightarrow\mathcal{C\ell}(TM)\otimes%
{\displaystyle\bigwedge\nolimits^{2}}
T^{\ast}M\hookrightarrow\mathcal{C\ell}(TM)\otimes\mathcal{C\ell}(T^{\ast
}M)\nonumber\\
\mathcal{J} &  =\mathbf{J_{\nu}\otimes}\theta^{\nu}\mathbf{\equiv J}_{\nu
}\theta^{\nu}\in\sec%
{\displaystyle\bigwedge\nolimits^{2}}
TM\otimes%
{\displaystyle\bigwedge\nolimits^{1}}
T^{\ast}M\hookrightarrow\mathcal{C\ell}(TM)\otimes\mathcal{C\ell}(T^{\ast
}M).\label{10.01}%
\end{align}

Now, using Eq.(\ref{a.hodge}) for the Hodge star operator given in the
Appendix A.3 and the relation between the operators $d=%
\mbox{\boldmath{$\partial$}}%
\wedge$ and $\delta=-%
\mbox{\boldmath{$\partial$}}%
\lrcorner$ \ (Appendix A5) we can write%
\begin{equation}
d\star\mathcal{R}=-\theta^{5}(-%
\mbox{\boldmath{$\partial$}}%
\lrcorner\mathcal{R)=-\star(}%
\mbox{\boldmath{$\partial$}}%
\lrcorner\mathcal{R)}=-\star((\partial_{\mu}\mathbf{R}_{\nu}^{\mu})\theta
^{\nu}). \label{10.02}%
\end{equation}

Also,%
\begin{align}
\lbrack%
\mbox{\boldmath{$\omega$}}%
,\mathbf{\star}\mathcal{R}]  &  =[%
\mbox{\boldmath{$\omega$}}%
_{\mu},\mathbf{R}_{\alpha\beta}]\otimes\theta^{\mu}\wedge\mathbf{\star(}%
\theta^{\alpha}\wedge\theta^{\beta}\mathbf{)}\nonumber\\
&  =-[%
\mbox{\boldmath{$\omega$}}%
_{\mu},\mathbf{R}_{\alpha\beta}]\otimes\theta^{\mu}\wedge\theta^{5}%
\mathbf{(}\theta^{\alpha}\wedge\theta^{\beta}\mathbf{)}\nonumber\\
&  =-\frac{1}{2}[%
\mbox{\boldmath{$\omega$}}%
_{\mu},\mathbf{R}_{\alpha\beta}]\otimes\{\theta^{\mu}\theta^{5}\mathbf{(}%
\theta^{\alpha}\wedge\theta^{\beta}\mathbf{)+}\theta^{5}\mathbf{\mathbf{(}%
}\theta^{\alpha}\mathbf{\wedge}\theta^{\beta}\mathbf{\mathbf{)}}\theta^{\mu
}\mathbf{\}}\nonumber\\
&  =\frac{\mathbf{\theta}^{5}}{2}[%
\mbox{\boldmath{$\omega$}}%
_{\mu},\mathbf{R}_{\alpha\beta}]\otimes\{\theta^{\mu}\mathbf{(}\theta^{\alpha
}\wedge\theta^{\beta}\mathbf{)-\mathbf{(}}\theta^{\alpha}\wedge\theta^{\beta
})\theta^{\mu}\mathbf{\}}\nonumber\\
&  =\mathbf{\theta}^{5}[%
\mbox{\boldmath{$\omega$}}%
_{\mu},\mathbf{R}_{\alpha\beta}]\otimes\{\theta^{\mu}\lrcorner\mathbf{(}%
\theta^{\alpha}\wedge\theta^{\beta}\mathbf{)}\nonumber\\
&  =-2\star([%
\mbox{\boldmath{$\omega$}}%
_{\mu},\mathbf{R}_{\beta}^{\mu}]\theta^{\beta}. \label{10.03}%
\end{align}

Using Eqs.(\ref{10.0}-\ref{10.03}) we get\footnote{Recall that $\mathbf{J}%
_{\nu}\in\sec\bigwedge\nolimits^{2}TM\hookrightarrow\sec\mathcal{C\ell
(}TM\mathcal{)}$.
\par
{}}
\begin{equation}
\mathcal{\partial}_{\mu}\mathbf{R}_{\nu}^{\mu}+2[%
\mbox{\boldmath{$\omega$}}%
_{\mu},\mathbf{R}_{\nu}^{\mu}]=\mathbf{D}_{e_{\mu}}\mathbf{R}_{\nu}^{\mu
}=\mathbf{J}_{\nu}.\label{10.1}%
\end{equation}
So, \ the gauge theory of gravitation has as field equations
the\ Eq.(\ref{10.1}), the nonhomogeneous field equations, and Eq. (\ref{9.28})
the homogeneous field equations (which is Bianchi's identity). We summarize
that equations, as
\begin{equation}
\mathbf{D}_{e_{\mu}}\mathbf{R}_{\nu}^{\mu}=\mathbf{J}_{\nu}\text{,\hspace{1cm}
}\mathbf{D}_{e_{\rho}}\mathbf{R}_{\mu\nu}+\mathbf{D}_{e_{\mu}}\mathbf{R}%
_{\nu\rho}+\mathbf{D}_{e_{\nu}}\mathbf{R}_{\rho\mu}=0.\label{10.1bis}%
\end{equation}

Eqs.(\ref{10.1bis}) which looks like Maxwell equations, must, of course, be
compatible with Einstein's equations, which may be eventually used to
determine $\mathbf{R}_{\nu}^{\mu},%
\mbox{\boldmath{$\omega$}}%
_{\mu}$ and $\mathbf{J}_{\nu}$.

\section{Another Set of Maxwell-Like Nonhomogeneous Equations for Einstein
Theory}

We now show, e.g., how a special combination of the $\mathbf{R}_{\mathbf{b}%
}^{\mathbf{a}}$ are directly related with a combination of products of the
energy-momentum $1$-vectors $T_{\mathbf{a}}$\ and the tetrad fields
$\mathbf{e}_{\mathbf{a}}$ (see Eq.(\ref{10.3bis}) below) in Einstein theory.
In order to do that, we recall that Einstein's equations can be written in
components in an orthonormal basis as
\begin{equation}
R_{\mathbf{ab}}-\frac{1}{2}\eta_{\mathbf{ab}}R=T_{\mathbf{ab}}, \label{10.2}%
\end{equation}
where $R_{\mathbf{ab}}=R_{\mathbf{ba}}$ are the components of the Ricci tensor
($R_{\mathbf{ab}}=R_{\mathbf{a}\text{ }\mathbf{bc}}^{\text{ }\mathbf{c}}$),
$T_{\mathbf{ab}}$ are the components of the energy-momentum tensor of matter
fields and $R=\eta_{\mathbf{ab}}R^{\mathbf{ab}}$ is the curvature scalar. We
next introduce\footnote{Ricci $1$-form fields appear naturally when we
formulate Einstein's equations in temrs of tetrad fields. See Appendix B.} the
\textit{Ricci 1-vectors} and the \textit{energy-momentum 1-vectors} by
\begin{align}
R_{\mathbf{a}}  &  =R_{\mathbf{ab}}\mathbf{e}^{\mathbf{b}}\in\sec%
{\displaystyle\bigwedge\nolimits^{1}}
TM\hookrightarrow\mathcal{C\ell}(TM),\label{10.3}\\
\hspace{0.15in}T_{\mathbf{a}}  &  =T_{\mathbf{ab}}\mathbf{e}^{\mathbf{b}}%
\in\sec%
{\displaystyle\bigwedge\nolimits^{1}}
TM\hookrightarrow\mathcal{C\ell}(TM). \label{10.3bis}%
\end{align}

We have that
\begin{equation}
R_{\mathbf{a}}=-\mathbf{e}^{\mathbf{b}}\lrcorner\mathbf{R}_{\mathbf{ab}}.
\label{10.4}%
\end{equation}

Now, multiplying Eq.(\ref{10.2}) on the right by $\mathbf{e}^{\mathbf{b}}$ we
get
\begin{equation}
R_{\mathbf{a}}-\frac{1}{2}R\mathbf{e}_{\mathbf{a}}=T_{\mathbf{a}}.
\label{10.7}%
\end{equation}

Multiplying Eq.(\ref{10.7}) first on the right by $\mathbf{e}_{\mathbf{b}}$
and then on the left by $\mathbf{e}_{\mathbf{b}}$ and making the difference of
the resulting equations we get%
\begin{equation}
\left(  -\mathbf{e}^{\mathbf{c}}\lrcorner\mathbf{R}_{\mathbf{ac}}\right)
\mathbf{e}_{\mathbf{b}}-\mathbf{e}_{\mathbf{b}}\left(  -\mathbf{e}%
^{\mathbf{c}}\lrcorner\mathbf{R}_{\mathbf{ac}}\right)  -\frac{1}%
{2}R(\mathbf{e}_{\mathbf{a}}\mathbf{e}_{\mathbf{b}}-\mathbf{e}_{\mathbf{b}%
}\mathbf{e}_{\mathbf{a}})=(T_{\mathbf{a}}\mathbf{e}_{\mathbf{b}}%
-\mathbf{e}_{\mathbf{b}}T_{\mathbf{a}}). \label{10.8}%
\end{equation}

Defining%

\begin{align}
\mathcal{F}_{\mathbf{ab}}  &  =\left(  -\mathbf{e}^{\mathbf{c}}\lrcorner
\mathbf{R}_{\mathbf{ac}}\right)  \mathbf{e}_{\mathbf{b}}-\mathbf{e}%
_{\mathbf{b}}\left(  -\mathbf{e}^{\mathbf{c}}\lrcorner\mathbf{R}_{\mathbf{ac}%
}\right)  -\frac{1}{2}R(\mathbf{e}_{\mathbf{a}}\mathbf{e}_{\mathbf{b}%
}-\mathbf{e}_{\mathbf{b}}\mathbf{e}_{\mathbf{a}})\nonumber\\
&  =\frac{1}{2}(R_{\mathbf{ac}}\mathbf{e}^{\mathbf{c}}\mathbf{e}_{\mathbf{b}%
}+\mathbf{e}_{\mathbf{b}}\mathbf{e}^{\mathbf{c}}R_{\mathbf{ac}}-\mathbf{e}%
^{\mathbf{c}}R_{\mathbf{ac}}\mathbf{e}_{\mathbf{b}}-\mathbf{e}_{\mathbf{b}%
}R_{\mathbf{ac}}\mathbf{e}^{\mathbf{c}})-\frac{1}{2}R(\mathbf{e}_{\mathbf{a}%
}\mathbf{e}_{\mathbf{b}}-\mathbf{e}_{\mathbf{b}}\mathbf{e}_{\mathbf{a}})
\label{10.9}%
\end{align}
and
\begin{equation}
\mathcal{J}_{\mathbf{b}}=D_{\mathbf{e}_{\mathbf{a}}}(T^{\mathbf{a}}%
\mathbf{e}_{\mathbf{b}}-\mathbf{e}_{\mathbf{b}}T^{\mathbf{a}}), \label{10.10}%
\end{equation}
we have\footnote{Note that we could also produce another Maxwell-like
equation, by using the extended covariant derivative operator in the
definition of the current, i.e., we can put $\mathcal{J}_{\mathbf{b}%
}=D_{\mathbf{e}_{\mathbf{a}}}(T^{\mathbf{a}}\mathbf{e}_{\mathbf{b}}%
-\mathbf{e}_{\mathbf{b}}T^{\mathbf{a}})$, and in that case we obtain
$\mathbf{D}_{\mathbf{e}_{\mathbf{a}}}\mathcal{F}_{\mathbf{b}}^{\mathbf{a}%
}=\mathcal{J}_{\mathbf{b}}$.}
\begin{equation}
D_{\mathbf{e}_{\mathbf{a}}}\mathcal{F}_{\mathbf{b}}^{\mathbf{a}}%
=\mathcal{J}_{\mathbf{b}}. \label{10.11}%
\end{equation}

It is quite obvious that in a coordinate chart $\langle x^{\mu}\rangle$
covering an open set $U\subset M$ we can write%
\begin{equation}
D_{e_{\rho}}\mathcal{F}_{\beta}^{\rho}=\mathcal{J}_{\beta}, \label{10.11bis}%
\end{equation}
with $\mathcal{F}_{\beta}^{\rho}=g^{\rho\alpha}\mathcal{F}_{\alpha\beta}$%
\begin{align}
\mathcal{F}_{\alpha\beta}  &  =\left(  -e^{\gamma}\lrcorner\mathbf{R}%
_{\alpha\gamma}\right)  e_{\beta}-e_{\beta}\left(  -e^{\gamma}\lrcorner
\mathbf{R}_{\alpha\gamma}\right)  -\frac{1}{2}R(e_{\alpha}e_{\beta}-e_{\beta
}e_{\alpha})\label{10.12}\\
\mathcal{J}_{\beta}  &  =D_{e_{\rho}}(T^{\rho}e_{\beta}-e^{\rho}T_{\beta}).
\label{10.12bis}%
\end{align}

\begin{remark}
\label{nat of F and J}Eq.(\ref{10.11}) (or Eq.(\ref{10.11bis})) is a set
Maxwell-like nonhomogeneous equations. It looks like the nonhomogeneous
classical Maxwell equations when that equations are written in components, but
Eq.(\ref{10.11bis}) is only a new way of writing the equation of the
nonhomogeneous field equations in the $Sl(2,\mathbb{C})$ like gauge theory
version of Einstein's theory, discussed in the previous section. In
particular, recall that any one of the six $\mathcal{F}_{\beta}^{\rho}\in\sec%
{\displaystyle\bigwedge\nolimits^{2}}
TM\hookrightarrow\mathcal{C}\ell(TM)$. Or, in words, each one of the
$\mathcal{F}_{\beta}^{\rho}$ is a bivector field, \textit{not} a set of
scalars which are components of a 2-form, as is the case in Maxwell theory.
Also, recall that according to Eq.(\ref{10.12bis}) each one of the four
$\mathcal{J}_{\beta}\in\sec\bigwedge\nolimits^{2}TM\hookrightarrow
\mathcal{C}\ell(TM)$.
\end{remark}

From Eq.(\ref{10.11}) \ it is not obvious that we must have $\mathcal{F}%
_{\mathbf{ab}}=0$ in vacuum, however that is exactly what happens if we take
into account Eq.(\ref{10.9}) which defines that object. Moreover,
$\mathcal{F}_{\mathbf{ab}}=0$\ does not imply that the curvature bivectors
$\mathbf{R}_{\mathbf{ab}}$ are null in vacuum. Indeed, in that case,
Eq.(\ref{10.9} ) implies only the identity (valid \textit{only} in vacuum)%

\begin{equation}
\left(  \mathbf{e}^{\mathbf{c}}\lrcorner\mathbf{R}_{\mathbf{ac}}\right)
\mathbf{e}_{\mathbf{b}}=\left(  \mathbf{e}^{\mathbf{c}}\lrcorner
\mathbf{R}_{\mathbf{bc}}\right)  \mathbf{e}_{\mathbf{a}}. \label{10.10''}%
\end{equation}

Moreover, recalling definition (Eq.(\ref{9.44})) we have%
\begin{equation}
\mathbf{R}_{\mathbf{ab}}=R_{\mathbf{abcd}}\mathbf{e}^{\mathbf{c}}%
\mathbf{e}^{\mathbf{d}}, \label{9.44'}%
\end{equation}
and we see that\ the $\mathbf{R}_{\mathbf{ab}}$ are zero only if the Riemann
tensor is null which is not the case in any non trivial general relativistic model.

The important fact that we want to emphasize here is that although eventually
interesting, Eq.(\ref{10.11}) does not seem (according to our opinion) to
contain anything new in it. More precisely, all information given by that
equation is already contained in the original Einstein's equation, for indeed
it has been obtained from it by simple algebraic manipulations. We state
again: According to our view terms like%

\begin{align}
\mathcal{F}_{\mathbf{ab}}  &  =\frac{1}{2}(R_{\mathbf{ac}}\mathbf{e}%
^{\mathbf{c}}\mathbf{e}_{\mathbf{b}}+\mathbf{e}_{\mathbf{b}}\mathbf{e}%
^{\mathbf{c}}R_{\mathbf{ac}}-\mathbf{e}^{\mathbf{c}}R_{\mathbf{ac}}%
\mathbf{e}_{\mathbf{b}}-\mathbf{e}_{\mathbf{b}}R_{\mathbf{ac}}\mathbf{e}%
^{\mathbf{c}})-\frac{1}{2}R(\mathbf{e}_{\mathbf{a}}\mathbf{e}_{\mathbf{b}%
}-\mathbf{e}_{\mathbf{b}}\mathbf{e}_{\mathbf{a}}),\nonumber\\
\mathfrak{R}_{\mathbf{ab}}  &  =(T_{\mathbf{a}}\mathbf{e}_{\mathbf{b}%
}-\mathbf{e}_{\mathbf{b}}T_{\mathbf{a}})-\frac{1}{2}R(\mathbf{e}_{\mathbf{a}%
}\mathbf{e}_{\mathbf{b}}-\mathbf{e}_{\mathbf{b}}\mathbf{e}_{\mathbf{a}%
}),\nonumber\\
\mathbf{F}_{\mathbf{ab}}  &  =\frac{1}{2}R(\mathbf{e}_{\mathbf{a}}%
\mathbf{e}_{\mathbf{b}}-\mathbf{e}_{\mathbf{b}}\mathbf{e}_{\mathbf{a}}),
\label{10.13}%
\end{align}
are pure gravitational objects. We cannot see any relationship of any one of
these objects with the ones appearing in Maxwell theory. Of course, these
objects may eventually be used to formulate interesting equations, like
Eq.(\ref{10.11}) which are equivalent to Einstein's field equations, but this
fact does not seem to us to point to any new Physics.\footnote{Note that
$\mathbf{F}_{\mathbf{ab}}$ differs from a factor, namely $R$ from the
$\mathbf{F}_{\mathbf{ab}}^{\prime}$ give by Eq.(70) in \cite{twospinor}.} Even
more, from the mathematical point of view, to find solutions to the new
Eq.(\ref{10.11}) is certainly as hard as to find solutions to the original
Einstein equations.

\subsection{$Sl\left(  2,\mathbb{C}\right)  $ Gauge Theory and Sachs
Antisymmetric Equation}

We discuss in this subsection yet another algebraic exercise. First recall
that in section 2 of \cite{twospinor} we define the paravector fields,%

\[
\mathbf{q}_{\mathbf{a}}=\mathbf{e}_{\mathbf{a}}\mathbf{e}_{\mathbf{0}}=%
\mbox{\boldmath{$\sigma$}}%
_{\mathbf{a}},\hspace{0.15in}\mathbf{\check{q}}_{\mathbf{a}}=(-%
\mbox{\boldmath{$\sigma$}}%
_{\mathbf{0}},%
\mbox{\boldmath{$\sigma$}}%
_{\mathbf{i}}),\hspace{0.15in}%
\mbox{\boldmath{$\sigma$}}%
_{\mathbf{0}}=1.
\]

Recall that\footnote{In Sachs book he wrote: $[D_{e_{\rho}},D_{e_{\lambda}%
}]e_{\mu}=R_{\mu\text{ }\rho\lambda}^{\text{ }\alpha}e_{\alpha}=+R_{\alpha
\mu\rho\lambda}e^{\alpha}$. This produces some changes in signals in relation
to our formulas below. Our Eq.(\ref{9.41}) agrees with the conventions in
\cite{choquet}.}
\begin{align}
\lbrack D_{e_{\rho}},D_{e_{\lambda}}]e_{\mu}  &  =R_{\mu\text{ }\rho\lambda
}^{\text{ }\alpha}e_{\alpha}=-R_{\alpha\mu\rho\lambda}e^{\alpha}=R_{\mu
\alpha\rho\lambda}e^{\alpha},\nonumber\\
R_{\mu\text{ }\rho\lambda}^{\text{ }\alpha}  &  =\mathcal{R}(e_{\mu}%
,\theta^{\alpha},e_{\rho},e_{\lambda}). \label{9.41}%
\end{align}

Then a simple calculation shows that
\begin{align}
\lbrack D_{e_{\rho}},D_{e_{\lambda}}]e_{\mu}  &  =e_{\mu}\lrcorner
\mathbf{R}_{\rho\lambda}=-\mathbf{R}_{\rho\lambda}\llcorner e_{\mu
},\label{9.43a}\\
R_{\mu\alpha\rho\lambda}e^{\alpha}  &  =\frac{1}{2}(e_{\mu}\mathbf{R}%
_{\rho\lambda}-\mathbf{R}_{\rho\lambda}e_{\mu}). \label{9.43b}%
\end{align}

Multiplying Eq.(\ref{9.43b}) on the left by $\mathbf{e}_{\mathbf{0}}$ we get,
recalling that $%
\mbox{\boldmath{$\omega$}}%
_{\mathbf{e}_{\mathbf{a}}}^{\dagger}=-\mathbf{e}^{\mathbf{0}}%
\mbox{\boldmath{$\omega$}}%
_{\mathbf{e}_{\mathbf{a}}}\mathbf{e}^{\mathbf{0}}$ (Eq.(79) in
\cite{twospinor}) we get%
\begin{equation}
R_{\mu\alpha\rho\lambda}\mathbf{q}^{\alpha}=\frac{1}{2}(\mathbf{q}_{\mu
}\mathbf{R}_{\rho\lambda}^{\dagger}+\mathbf{R}_{\rho\lambda}\mathbf{q}_{\mu}).
\label{9.43c}%
\end{equation}
\ 

Now, to derive Sachs\footnote{Numeration is from Sachs' book \cite{s1}.}
Eq.(6.50a) all we need to do is to multiply Eq.(\ref{10.8}) on the right by
$\mathbf{e}^{\mathbf{0}}$ and perform some algebraic manipulations. We then
get (with \textit{our} normalization) for the equivalent of Einstein's
equations using the paravector fields and a coordinate chart $\langle x^{\mu
}\rangle$ covering an open set $U\subset M$, the following equation
\begin{equation}
\mathbf{R}_{\rho\lambda}\mathbf{q}^{\lambda}+\mathbf{q}^{\lambda}%
\mathbf{R}_{\rho\lambda}^{\dagger}+R\mathbf{q}_{\rho}=2\mathbf{T}_{\rho}.
\label{sachs1}%
\end{equation}
For the Hermitian conjugate we have
\begin{equation}
-\mathbf{R}_{\rho\lambda}^{\dagger}\mathbf{\check{q}}^{\lambda}-\mathbf{\check
{q}}^{\lambda}\mathbf{R}_{\rho\lambda}+R\mathbf{\check{q}}_{\rho
}=2\mathbf{\check{T}}_{\rho}, \label{sachs1'}%
\end{equation}
where as above, the $\mathbf{R}_{\rho\lambda}$ are the curvature bivectors
given by Eq.(\ref{9.40}) and
\begin{equation}
\mathbf{T}_{\rho}=T_{\rho}^{\mu}\mathbf{q}_{\mu}\in\sec\bigwedge
\nolimits^{2}TM\hookrightarrow\mathcal{C\ell(}TM\mathcal{)}. \label{sachs2}%
\end{equation}

After that, we multiply Eq.(\ref{sachs1}) on the right by $\mathbf{\check{q}%
}_{\gamma}$ and Eq.(\ref{sachs1'}) on the left by $\mathbf{q}_{\gamma}$ ending
with two new equations. If we sum them, we get a \ `symmetric'
equation\footnote{Eq.(6.52) in Sachs' book \cite{s1}.} completely equivalent
to Einstein's equation (from where we started). If we make the difference of
the equations we get an antisymmetric equation. The antisymmetric equation can
be written, introducing%
\begin{align}
\mathbb{F}_{\rho\gamma}  &  =\frac{1}{2}(\mathbf{R}_{\rho\lambda}%
\mathbf{q}^{\lambda}\mathbf{\check{q}}_{\gamma}\mathbf{+q}_{\gamma
}\mathbf{\check{q}}^{\lambda}\mathbf{R}_{\rho\lambda}\mathbf{+q}^{\lambda
}\mathbf{R}_{\rho\lambda}^{\dagger}\mathbf{\check{q}}_{\gamma}\mathbf{+q}%
_{\gamma}\mathbf{R}_{\rho\lambda}^{\dagger}\mathbf{\check{q}}^{\lambda
})\label{sachs3}\\
&  +\frac{1}{2}R(\mathbf{q}_{\rho}\mathbf{\check{q}}_{\gamma}\mathbf{-q}%
_{\gamma}\mathbf{\check{q}}_{\rho})\nonumber
\end{align}
and%
\begin{equation}
\mathbb{J}_{\gamma}=D_{e_{\rho}}(\mathbf{T}^{\rho}\mathbf{\check{q}}_{\gamma
}-\mathbf{q}_{\gamma}\mathbf{\check{T}}^{\rho}), \label{sachs4}%
\end{equation}
as%
\begin{equation}
D_{e_{\rho}}\mathbb{F}_{\gamma}^{\rho}=\mathbb{J}_{\gamma}. \label{sachs5}%
\end{equation}
\ \ 

\begin{remark}
It is important to keep in mind that each one of the six $\mathbb{F}%
_{\rho\gamma}$ and each one of the four $\mathbb{J}_{\gamma}$ are not a set of
scalars, but sections of $\mathcal{C}\ell^{(0)}\mathcal{(}TM\mathcal{)}$.
Also, take notice that Eq.(\ref{sachs5}), of course, is completely equivalent
to our Eq.(\ref{10.11}). Its matrix translation in $\mathbb{C}\ell
^{(0)}(M)\simeq S(M)\otimes_{\mathbb{C}}\bar{S}(M)$ gives Sachs equation
(6.52-) in \cite{s1} if we take into account his different \ `normalization'
of the connection coefficients and the \textit{ad hoc} factor with dimension
of electric charge that he introduced. We cannot see at present any new
information encoded in that equations which could be translated in
\ interesting geometrical properties of the manifold, but of course,
eventually someone may find that they encode such a useful
information.\footnote{Anyway, it seems to us that until the written of the
present paper the true mathematical nature of Sachs equations have not been
understood, by people that read Sachs books and articles. To endorse our
statement, we quote that in Carmeli's review(\cite{carmelimr}) of Sachs book,
he did not realize that Sachs theory was indeed (as we showed above) a
description in the Pauli bundle of a $Sl(2,\mathbb{C)}$ gauge formulation of
Einstein's theory as described in his own book \cite{carmeli}. Had he
disclosed that fact (as we did) he probably had not written that Sachs'
approach was a possible unified field theory of gravitation and
electromagnetism.}
\end{remark}

Using \ the equations, \ $D_{\mathbf{e}_{\mathbf{a}}}\mathbf{e}_{0}=0$ and
$D_{e_{\rho}}^{\mathbf{S}}\mathbf{q}_{\mu}=0$ (respectively, \ Eq.(88) and
Eq.(108) in \cite{twospinor} ) and (\ref{9.28}) we may verify that
\begin{equation}
D_{e_{\rho}}^{\mathbf{S}}\mathbb{F}_{\mu\nu}+D_{e_{\mu}}^{\mathbf{S}%
}\mathbb{F}_{\nu\rho}+D_{e_{\nu}}^{\mathbf{S}}\mathbb{F}_{\rho\mu
}=0,\label{sachs6}%
\end{equation}
where $D_{e_{\rho}}^{\mathbf{S}}$ is Sachs `covariant' derivative that we
discussed in \cite{twospinor}. In \cite{s2} Sachs concludes that the last
equation implies that there are no magnetic monopoles in nature. Of course,
his conclusion would follow from Eq.(\ref{sachs6}) only if it happened that
$\mathbb{F}_{\gamma}^{\rho}$ were the components in a coordinate basis of a
$2$-form field $F\in\sec\bigwedge\nolimits^{2}T^{\ast}M$. However, this is not
the case, because as already noted above, this is not the mathematical nature
of the $\mathbb{F}_{\gamma}^{\rho}$ . Contrary to what we stated with relation
to Eq.(\ref{sachs5}) we cannot even say that Eq.(\ref{sachs6}) is really
interesting, because it uses a covariant derivative operator, which, as
discussed in \cite{twospinor} is not \ well justified, and in anyway
$D_{e_{\rho}}^{\mathbf{S}}\neq D_{e_{\rho}}$. We cannot see any relationship
of Eq.(\ref{sachs6}) with the legendary magnetic monopoles.

We thus conclude this section stating that Sachs claims in \cite{s1,s2,s3} of
having produced an unified field theory of electricity and electromagnetism
are not endorsed by our analysis.

\section{Energy-Momentum \textquotedblleft Conservation\textquotedblright\ in
General Relativity}

\subsection{Einstein's Equations in terms of Superpotentials $\star
S^{\mathbf{a}}$}

In this section we discuss some issues and statements concerning the problem
of the energy-momentum conservation in Einstein's theory, presented with
several different formalisms in the literature, which according to our view
are very confusing, or even wrong. To start, recall that from Eq.(\ref{10.0})
it follows that
\begin{equation}
d(\star\mathcal{J}\mathbf{-}\frac{1}{2}[%
\mbox{\boldmath{$\omega$}}%
,\mathbf{\star}\mathcal{R}])=0, \label{10.0BISS}%
\end{equation}
and we could think that this equation could be used to identify a
\textit{conservation} \textit{law} for the energy momentum of matter plus the
gravitational field, with $\frac{1}{2}[%
\mbox{\boldmath{$\omega$}}%
,\mathbf{\star}\mathcal{R}]$ describing a mathematical object related to the
energy- momentum of the gravitational field. However, this is not the case,
because this term (due to the presence of $%
\mbox{\boldmath{$\omega$}}%
$) is gauge dependent. The appearance of a gauge dependent term is \ a
recurrent fact in all known proposed\footnote{At least, the ones known by the
authors.
\par
{}} formulations of a \ `conservation law for energy-momentum' for Einstein
theory. We discuss now some statements found in the literature based on some
of that proposed \ `solutions' to the problem of energy-momentum conservation
in General Relativity and say why we think they are unsatisfactory. We also
mention a way with which the problem could be satisfactorily solved, but which
implies in a departure from the orthodox interpretation of Einstein's theory.

Now, to keep the mathematics as simple and transparent as possible, instead of
working with Eq.(\ref{10.0BISS}), we work with a more simple (but equivalent)
formulation \cite{rq,thirring} of Einstein's equation where the gravitational
field is described by a set of $2$-forms $\star S^{\mathbf{a}}$,
$\mathbf{a}=0,1,2,3$ \ called superpotentials. This approach will permit to
identify very quickly certain objects that at first sight seems
\textit{appropriate \ energy-momentum }currents for the gravitational field in
Einstein's theory. The calculations that follows are done in the Clifford
algebra of multiforms fields $\mathcal{C\ell}\left(  T^{\ast}M\right)  $,
something that, as the reader will testify, simplify considerably similar
calculations done with traditional methods.

We start again with Einstein's equations given by Eq.(\ref{10.2}), but this
time we multiply on the left by $\theta^{\mathbf{b}}\in\sec\bigwedge
\nolimits^{1}T^{\ast}M\hookrightarrow\mathcal{C\ell}\left(  T^{\ast}M\right)
$ getting an equation relating the \textit{Ricci }$\mathit{1}$\textit{-forms}
$\ \mathcal{R}^{\mathbf{a}}=R_{\mathbf{b}}^{\mathbf{a}}\theta^{\mathbf{b}}$
$\in\sec\bigwedge\nolimits^{1}T^{\ast}M\hookrightarrow\mathcal{C\ell}\left(
T^{\ast}M\right)  $ with the \textit{energy-momentum 1-forms} $\mathcal{T}%
^{\mathbf{a}}=T_{\mathbf{b}}^{\mathbf{a}}\theta^{\mathbf{b}}\in\sec
\bigwedge\nolimits^{1}T^{\ast}M\hookrightarrow\mathcal{C\ell}\left(  T^{\ast
}M\right)  $, i.e.,
\begin{equation}
\mathcal{G}^{\mathbf{a}}=\mathcal{R}^{\mathbf{a}}-\frac{1}{2}R\theta
^{\mathbf{a}}=\mathcal{T}^{\mathbf{a}}. \label{10.14}%
\end{equation}

We take the dual of this equation,%
\begin{equation}
\star\mathcal{G}^{\mathbf{a}}=\star\mathcal{T}^{\mathbf{a}}. \label{10.15}%
\end{equation}

Next, we observe that \cite{rq,thirring} we can write%
\begin{equation}
\star\mathcal{G}^{\mathbf{a}}=-d\star\mathcal{S}^{\mathbf{a}}-\star
\mathfrak{t}_{\mathbf{\ }}^{\mathbf{a}}, \label{10.16}%
\end{equation}
where \
\begin{align}
\star\mathcal{S}^{\mathbf{c}}  &  =\frac{1}{2}%
\mbox{\boldmath{$\omega$}}%
_{\mathbf{ab}}\wedge\star(\theta^{\mathbf{a}}\wedge\theta^{\mathbf{b}}%
\wedge\theta^{\mathbf{c}}),\nonumber\\
\star\mathfrak{t}_{\mathbf{\ }}^{\mathbf{c}}  &  =-\frac{1}{2}%
\mbox{\boldmath{$\omega$}}%
_{\mathbf{ab}}\wedge\lbrack%
\mbox{\boldmath{$\omega$}}%
_{\mathbf{d}}^{\mathbf{c}}\star(\theta^{\mathbf{a}}\wedge\theta^{\mathbf{b}%
}\wedge\theta^{\mathbf{d}})-%
\mbox{\boldmath{$\omega$}}%
_{\mathbf{d}}^{\mathbf{b}}\star(\theta^{\mathbf{a}}\wedge\theta^{\mathbf{d}%
}\wedge\theta^{\mathbf{c}})]. \label{10.17}%
\end{align}

The proof of Eq.(\ref{10.17}) follows at once from the fact that
\begin{equation}
\star\mathcal{G}^{\mathbf{d}}=-\frac{1}{2}\mathcal{R}_{\mathbf{ab}}\wedge
\star(\theta^{\mathbf{a}}\wedge\theta^{\mathbf{b}}\wedge\theta^{\mathbf{d}}).
\label{10.18}%
\end{equation}

Indeed, recalling the identities in Eq.(\ref{Aidentities}) we can write%
\begin{align}
\frac{1}{2}\mathcal{R}_{\mathbf{ab}}\wedge\star(\theta^{\mathbf{a}}%
\wedge\theta^{\mathbf{b}}\wedge\theta^{\mathbf{d}})  &  =-\frac{1}{2}%
\star\lbrack\mathcal{R}_{\mathbf{ab}}\lrcorner(\theta^{\mathbf{a}}\wedge
\theta^{\mathbf{b}}\wedge\theta^{\mathbf{d}})]\nonumber\\
&  =-\frac{1}{2}R_{\mathbf{abcd}}\star\lbrack(\theta^{\mathbf{c}}\wedge
\theta^{\mathbf{d}})\lrcorner(\theta^{\mathbf{a}}\wedge\theta^{\mathbf{b}%
}\wedge\theta^{\mathbf{d}})]\nonumber\\
&  =-\star(\mathcal{R}^{\mathbf{d}}-\frac{1}{2}R\theta^{\mathbf{d}}).
\label{10.19}%
\end{align}

On the other hand we have,%
\begin{align}
&  -2\star\mathcal{G}^{\mathbf{d}}=d%
\mbox{\boldmath{$\omega$}}%
_{\mathbf{ab}}\wedge\star(\theta^{\mathbf{a}}\wedge\theta^{\mathbf{b}}%
\wedge\theta^{\mathbf{d}})+%
\mbox{\boldmath{$\omega$}}%
_{\mathbf{ac}}\wedge%
\mbox{\boldmath{$\omega$}}%
_{\mathbf{b}}^{\mathbf{c}}\wedge\star(\theta^{\mathbf{a}}\wedge\theta
^{\mathbf{b}}\wedge\theta^{\mathbf{d}})\nonumber\\
&  =d[%
\mbox{\boldmath{$\omega$}}%
_{\mathbf{ab}}\wedge\star(\theta^{\mathbf{a}}\wedge\theta^{\mathbf{b}}%
\wedge\theta^{\mathbf{d}})]+%
\mbox{\boldmath{$\omega$}}%
_{\mathbf{ab}}\wedge d\star(\theta^{\mathbf{a}}\wedge\theta^{\mathbf{b}}%
\wedge\theta^{\mathbf{d}})\nonumber\\
&  +%
\mbox{\boldmath{$\omega$}}%
_{\mathbf{ac}}\wedge%
\mbox{\boldmath{$\omega$}}%
_{\mathbf{b}}^{\mathbf{c}}\wedge\star(\theta^{\mathbf{a}}\wedge\theta
^{\mathbf{b}}\wedge\theta^{\mathbf{d}})\nonumber\\
&  =d[%
\mbox{\boldmath{$\omega$}}%
_{\mathbf{ab}}\wedge\star(\theta^{\mathbf{a}}\wedge\theta^{\mathbf{b}}%
\wedge\theta^{\mathbf{d}})]-%
\mbox{\boldmath{$\omega$}}%
_{\mathbf{ab}}\wedge%
\mbox{\boldmath{$\omega$}}%
_{\mathbf{p}}^{\mathbf{a}}\wedge\star(\theta^{\mathbf{p}}\wedge\theta
^{\mathbf{b}}\wedge\theta^{\mathbf{d}})\nonumber\\
&  -%
\mbox{\boldmath{$\omega$}}%
_{\mathbf{ab}}\wedge%
\mbox{\boldmath{$\omega$}}%
_{\mathbf{p}}^{\mathbf{b}}\wedge\star(\theta^{\mathbf{a}}\wedge\theta
^{\mathbf{p}}\wedge\theta^{\mathbf{d}})-%
\mbox{\boldmath{$\omega$}}%
_{\mathbf{ab}}\wedge%
\mbox{\boldmath{$\omega$}}%
_{\mathbf{p}}^{\mathbf{d}}\wedge\star(\theta^{\mathbf{a}}\wedge\theta
^{\mathbf{b}}\wedge\theta^{\mathbf{p}})]\nonumber\\
&  +%
\mbox{\boldmath{$\omega$}}%
_{\mathbf{ac}}\wedge%
\mbox{\boldmath{$\omega$}}%
_{\mathbf{b}}^{\mathbf{c}}\wedge\star(\theta^{\mathbf{a}}\wedge\theta
^{\mathbf{b}}\wedge\theta^{\mathbf{d}})\nonumber\\
&  =d[%
\mbox{\boldmath{$\omega$}}%
_{\mathbf{ab}}\wedge\star(\theta^{\mathbf{a}}\wedge\theta^{\mathbf{b}}%
\wedge\theta^{\mathbf{d}})]-%
\mbox{\boldmath{$\omega$}}%
_{\mathbf{ab}}\wedge\lbrack%
\mbox{\boldmath{$\omega$}}%
_{\mathbf{p}}^{\mathbf{d}}\wedge\star(\theta^{\mathbf{a}}\wedge\theta
^{\mathbf{b}}\wedge\theta^{\mathbf{p}})+%
\mbox{\boldmath{$\omega$}}%
_{\mathbf{p}}^{\mathbf{b}}\wedge\star(\theta^{\mathbf{a}}\wedge\theta
^{\mathbf{p}}\wedge\theta^{\mathbf{d}})]\nonumber\\
&  =2(d\star\mathcal{S}^{\mathbf{d}}+\star\mathfrak{t}^{\mathbf{d}%
}).\label{10.20}%
\end{align}

Now, we can then write Einstein's equation in a very interesting, but
\textit{dangerous} form, i.e.,%
\begin{equation}
-d\star\mathcal{S}^{\mathbf{a}}=\star\mathcal{T}^{\mathbf{a}}+\star
\mathfrak{t}^{\mathbf{a}}. \label{10.21}%
\end{equation}

In writing Einstein's equations in that way, we have associated to the
gravitational field a set of $2$-form fields $\star\mathcal{S}^{\mathbf{a}}$
called \textit{superpotentials} that have as sources the currents
$(\star\mathcal{T}^{\mathbf{a}}+\star\mathfrak{t}^{\mathbf{a}})$. However,
superpotentials are not uniquely defined since, e.g., superpotentials
\ $(\star\mathcal{S}^{\mathbf{a}}+\star\alpha^{\mathbf{a}})$, with
$\star\alpha^{\mathbf{a}}$ closed, i.e., $d\star\alpha^{\mathbf{a}}=0$ give
the same second member for Eq.(\ref{10.21}).

\subsection{Is There Any Energy-Momentum Conservation Law in GR?}

Why did we say that Eq.(\ref{10.21}) is a dangerous one?

The reason is that (as in the case of Eq.(\ref{10.0BISS})) we can be led to
think that we have discovered a conservation law for the energy-momentum of
matter plus gravitational field, since from Eq.(\ref{10.21}) it follows that
\begin{equation}
d(\star\mathcal{T}^{\mathbf{a}}+\star\mathfrak{t}^{\mathbf{a}})=0.
\label{10.22}%
\end{equation}
This thought however is only an example of wishful thinking, because the
$\star\mathfrak{t}^{\mathbf{a}}$ \ depends on the connection (see
Eq.(\ref{10.17})) and thus are gauge dependent. They do not have the same
tensor transformation law as the $\star\mathcal{T}^{\mathbf{a}}$. So, Stokes
theorem cannot be used to derive from Eq.(\ref{10.22}) conserved quantities
that are independent of the gauge, which is clear. However, and this is less
known, Stokes theorem, also cannot be used to derive conclusions that are
independent of the local coordinate chart used to perform calculations
\cite{boro}. In fact, the currents $\star\mathfrak{t}^{\mathbf{a}}$ are
nothing more than the old pseudo energy momentum tensor of Einstein in a new
dress. Nonrecognition of this fact can lead to many misunderstandings. We
present some of them in what follows, in order to call our readers' attention
of potential errors of inference that can be done when we use sophisticated
mathematical formalisms without a perfect domain of their contents.

\qquad\textbf{(i)}\ First, it is easy to see that from Eq.(\ref{10.15}) it
follows that \cite{mtw}
\begin{equation}
\mathbf{D}^{c}\mathbf{\star}\mathfrak{G}=\mathbf{D}^{c}\star\mathfrak{T}=0,
\label{10.22.0}%
\end{equation}
where $\mathbf{\star}\mathfrak{G}=\mathbf{e}_{\mathbf{a}}\otimes
\star\mathcal{G}^{\mathbf{a}}$ $\in\sec TM\otimes\sec\bigwedge\nolimits^{3}%
T^{\ast}M$ and $\star\mathfrak{T}=\mathbf{e}_{\mathbf{a}}\otimes
\star\mathcal{T}^{\mathbf{a}}\in\sec TM\otimes\sec\bigwedge\nolimits^{3}%
T^{\ast}M$ . Now, in \cite{mtw} it is written (without proof) a \ `Stokes
theorem' \medskip%
\begin{equation}%
\begin{tabular}
[c]{|c|}\hline
$%
{\displaystyle\int\limits_{{\footnotesize 4}\text{-cube}}}
\mathbf{D}^{c}\mathbf{\star}\mathfrak{T}\mathbf{=}%
{\displaystyle\int\limits_{\substack{{\footnotesize 3}\text{ boundary}\\\text{
of this }{\footnotesize 4}\text{-cube}}}}
\mathbf{\star}\mathfrak{T}$\\\hline
\end{tabular}
\ \ \ \label{mtw}%
\end{equation}

\begin{center}
\medskip
\end{center}

We searched in the literature for a proof of Eq.(\ref{mtw}) which appears also
in many other texts and scientific papers, as e.g., in \cite{dalton,vatorr1}
and could find none, which we can consider as valid. The reason is simply. If
expressed in details, e.g., the first member of Eq.(\ref{mtw}) reads
\begin{equation}%
{\displaystyle\int\limits_{{\footnotesize 4}\text{-cube}}}
\mathbf{e}_{\mathbf{a}}\otimes(d\star\mathcal{T}^{\mathbf{a}}+\omega
_{\mathbf{b}}^{\mathbf{a}}\wedge\mathcal{T}^{\mathbf{b}}), \label{10.22.01}%
\end{equation}
and it is necessary to explain what is the meaning (if any) of the integral.
Since the integrand is a sum of tensor fields, this integral says that we are
\textit{summing} tensors belonging to the tensor spaces of different spacetime
points. As, well known, this cannot be done in general, unless there is a way
for identification of the tensor spaces at different spacetime points. This
requires, of course, the introduction of additional structure on the spacetime
representing a given gravitational field, and such extra structure is lacking
in Einstein's theory. We unfortunately, must conclude that Eq.(\ref{mtw}) do
not express any conservation law, for it lacks as yet, a precise mathematical
meaning.\footnote{Of course, if some could give a mathematical meaning to
Eq.(\ref{mtw}), we will be glad to be informed of that fact.}

\ \ In Einstein theory possible superpotentias are, of course, the
$\star\mathcal{S}^{\mathbf{a}}$ that we found above (Eq.(\ref{10.17})), with
\begin{equation}
\star\mathcal{S}_{\mathbf{c}}=[-\frac{1}{2}\omega_{\mathbf{ab}}\lrcorner
(\theta^{\mathbf{a}}\wedge\theta^{\mathbf{b}}\wedge\theta_{\mathbf{c}}%
)]\theta^{\mathbf{5}}. \label{10.22.1.1}%
\end{equation}

Then, if we integrate Eq.(\ref{10.21}) over a \ `certain finite $3$%
-dimensional volume', say a ball $B$, and use Stokes theorem we have%
\begin{equation}
P^{\mathbf{a}}=%
{\displaystyle\int\limits_{B}}
\star\left(  \mathcal{T}^{\mathbf{a}}+\mathfrak{t}^{\mathbf{a}}\right)  =-%
{\displaystyle\int\limits_{\partial B}}
\star\mathcal{S}^{\mathbf{a}}. \label{10.22.2}%
\end{equation}

In particular the energy or (\textit{inertial mass}) of the gravitational
field plus matter generating the field is defined by
\begin{equation}
P^{\mathbf{0}}=E=m_{i}=-\lim_{R\rightarrow\infty}%
{\displaystyle\int\limits_{\partial B}}
\star\mathcal{S}^{\mathbf{0}} \label{10.22.3'}%
\end{equation}

\textbf{(ii) }Now, a frequent misunderstanding is the following. Suppose that
in a \textit{given} gravitational theory\ there exists an energy-momentum
conservation law for matter plus the gravitational field expressed in the form
of Eq.(\ref{10.22}), where $\mathcal{T}^{\mathbf{a}}$ \ are the
energy-momentum 1-forms of matter and $\mathfrak{t}^{\mathbf{a}}$ are
\textit{true}\footnote{This means that the $t^{\mathbf{a}}$ are not pseudo
1-forms, as in Einstein's theory.} energy-momentum 1-forms of the
gravitational field. \ This means that \ the $3$-forms $(\mathbf{\star
}\mathcal{T}^{\mathbf{a}}+\star\mathfrak{t}^{\mathbf{a}})$ are closed, i.e.,
\ they satisfy Eq.(\ref{10.22}). Is this enough to warrant that the energy of
a closed universe is zero? Well, that would be the case if starting from
\ Eq.(\ref{10.22}) we could jump to an equation like Eq.(\ref{10.21}) and then
to Eq.(\ref{10.22.3'}) (as done, e.g., in \cite{thirring2}). But that sequence
of inferences in general cannot be done, for indeed, as it is well known, it
is not the case that closed three forms are always exact. Take a closed
universe with topology, say $\mathbb{R\times}S^{3}$. In this case $B=$ $S^{3}$
and we have $\partial B=$ $\partial S^{3}=\varnothing$. Now, as it is well
known (see, e.g., \cite{nakahara}), the third de Rham cohomology group of
$\mathbb{R\times}S^{3}$ is $H^{3}\left(  \mathbb{R\times}S^{3}\right)
=H^{3}\left(  S^{3}\right)  =\mathbb{R}$. Since this group is non trivial it
follows that in such manifold closed forms are not exact. Then from
Eq.(\ref{10.22}) it did not follow the validity of an equation analogous to
Eq.(\ref{10.21}). So, in that case an equation like Eq.(\ref{10.22.2}) cannot
even be written.

Despite that commentary, keep in mind that in Einstein's theory the energy of
a closed universe\footnote{Note that if we suppose that the universe contains
spinor fields, then it must be a spin manifold, i.e., it is parallezible
according to Geroch's theorem \cite{geroch}.}, if it is \ justifiable to
suppose that it is given by Eq.(\ref{10.22.3'}), is indeed zero, since in that
theory the $3$-forms $(\mathbf{\star}\mathcal{T}^{\mathbf{a}}+\star
\mathfrak{t}^{\mathbf{a}})$ are indeed exact (see Eq.(\ref{10.21})). This
means that accepting $\mathfrak{t}^{\mathbf{a}}$ as the \ `energy-momentum'
$1$-form fields of the gravitational field, it follows that
gravitational\ `energy' \textit{must} be negative in a closed universe.

\textbf{(iii)} But, is the above formalism a consistent one? Given a
coordinate chart $\langle x^{\mu}\rangle$ of the maximal atlas of $M$, with
some algebra we can show that for a gravitational model represented by a
diagonal asymptotic flat metric\footnote{A metric is said to be asymptotically
flat in given coordinates, if $g_{\mu\nu}=n_{\mu\nu}(1+\mathrm{O}\left(
r^{-k}\right)  )$, with $k=2$ or $k=1$ depending on the author. See, eg.,
\cite{schoenyau1, schoenyau2,wald}.}, the inertial mass $E=m_{i}$ is given by
\begin{equation}
m_{i}=\lim_{R\rightarrow\infty}\frac{-1}{16\pi}%
{\displaystyle\int\limits_{\partial B}}
\frac{\partial}{\partial x^{\beta}}(g_{11}g_{22}g_{33}g^{\alpha\beta}%
)d\sigma_{\alpha},\label{10.22.4}%
\end{equation}
where $\partial B=S^{2}(R)$ is a $2$-sphere of radius $R$,
$(-n_{\mathbf{\alpha}})$ is the outward unit normal and $d\sigma_{\alpha
}=-R^{2}n_{\alpha}dA$. \ If we apply Eq.(\ref{10.22.4}) to calculate, e.g.,
the energy of the Schwarzschild spacetime\footnote{For a Scharzschild
spacetime we have $g=\left(  1-\frac{2m}{r}\right)  dt\otimes dt-\left(
1-\frac{2m}{r}\right)  ^{-1}dr\otimes dr-r^{2}(d\theta\otimes d\theta+\sin
^{2}\theta d\varphi\otimes d\varphi)$.} generate by a gravitational mass $m$,
we expect to have one unique and unambiguous result, namely $m_{i}=m$.

However, as showed in details, e.g., in \cite{boro} the calculation of $E$
depends on the spatial coordinate system naturally adapted to the reference
frame $Z=\frac{1}{\sqrt{\left(  1-\frac{2m}{r}\right)  }}\frac{\partial
}{\partial t}$ , even if these coordinates produce asymptotically flat
metrics. Then, even if in one given chart we may obtain as the result of the
calculation $m_{i}=m$ there are others where the results of the calculations
give $m_{i}\neq m$!

Moreover, note also that, as showed above, for a closed universe, Einstein's
theory implies on general grounds (once we accept that the $\mathfrak{t}^{a}$
describes the energy-momentum distribution of the gravitational field) that
$m_{i}=0$. This result, it is important to quote, does not contradict the so
called "positive mass theorems" of, e.g., references
\cite{schoenyau1,schoenyau2,witten}, because that theorems refers to the total
\ `energy' of an isolated system. A system of that kind is supposed to be
modelled by a Lorentzian spacetime having a spacelike, asymptotically
Euclidean hypersurface.\footnote{The proof also uses as hypothesis the so
called energy dominance condition. \cite{hawellis}} However, we want to
emphasize here, that although the \ `energy' results positive, its value is
not unique, since depends on the asymptotically flat coordinates chosen to
perform the calculations, as it is clear from the example of the Schwarzschild
\ field, as we already commented above and detailed in \cite{boro}.

In view of what has been presented above, it is our view that all discourses
(based on Einstein's equivalence principle) concerning the use of
pseudo-energy momentum tensors as \textit{reasonable} descriptions of energy
and momentum of gravitational fields in Einstein's theory are not convincing.

The fact is: there are \textit{in general} no conservation laws of
energy-momentum in General Relativity in general. And, at this point it is
better to quote page 98 of Sachs\&Wu\footnote{Note, please, that in this
reference Sachs refers to R. K. Sachs and not to M. Sachs.} \cite{sw}:

{\footnotesize \ " As mentioned in section 3.8, conservation laws have a great
predictive power. It is a shame to lose the special relativistic total energy
conservation law (Section 3.10.2) in general relativity. Many of the attempts
to resurrect it are quite interesting; many are simply garbage."}

We quote also Anderson \cite{anderson}:

{\footnotesize " In an interaction that involves the gravitational field a
system can loose energy without this energy being transmitted to the
gravitational field."}

In General Relativity, we already said, every gravitational field is modelled
(module diffeomorphisms) by a Lorentzian spacetime. In the particular case,
when this spacetime structure admits a \textit{timelike} Killing vector, we
can formulate a law of energy conservation. If the spacetime admits three
linearly independent \textit{spacelike} Killing vectors, we have a law of
conservation of momentum. The crucial fact to have in mind here is that a
general Lorentzian spacetime, does not admits such Killing vectors in general.
As one example, we quote that the popular Friedmann-Robertson-Walker expanding
universes models do not admit timelike Killing vectors, in general.

At present, the authors know only one possibility of resurrecting a
\textit{trustworthy} conservation law of energy-momentum valid in all
circumstances in a theory of the gravitational field that \textit{resembles}
General Relativity (in the sense of keeping Einstein's equation). It consists
in reinterpreting that theory as a field theory in flat Minkowski spacetime.
Theories of this kind have been proposed in the past by, e.g., Feynman
\cite{feynman}, Schwinger \cite{schwinger},Thirring \cite{thirring0} and
Weinberg \cite{weinberg1,weinberg2} and have been extensively studied by
Logunov and collaborators \cite{logunov1,logunov2}. Another presentation of a
theory of that kind, is one where the gravitational field is represented by a
distortion field in Minkowski spacetime. A first attempt to such a theory
using Clifford bundles has been given in \cite{rq}. Another presentation has
been given in \cite{ladogu}, but that work, which contains many interesting
ideas, unfortunately contains also some equivocated statements that make (in
our opinion) the theory, as originally presented by that authors invalid. This
has been discussed with details in \cite{femoro}.

Before closing this section we observe that recently people think to have
found a valid way of having a genuine energy-momentum conservation law in
general relativity, by using the so-called \textit{teleparallel} version of
that theory \cite{deandrade}. If that is really the case will be analyzed in a
sequel paper \cite{roqui}, where we discuss conservation laws in a general
Riemann-Cartan spacetime, using Clifford bundle methods.

\section{Conclusions}

In this paper we introduced the concept of Clifford valued differential forms,
which are sections of $\mathcal{C\ell}(TM)\otimes%
{\displaystyle\bigwedge}
T^{\ast}M$. We showed how this theory can be used to produce a very elegant
description of the theory of linear connections, where a given linear
connection is represented by a bivector valued 1-form. Crucial to the program
was the introduction of the notion of the exterior covariant differential and
the extended covariant derivative acting on sections of $\mathcal{C\ell
}(TM)\otimes%
{\displaystyle\bigwedge}
T^{\ast}M$. Our \textit{natural} definitions parallel in a noticeable way the
formalism of the theory of connections in a principal bundle and the covariant
derivative operators acting on associate bundles to that principal bundle. We
identified Cartan curvature 2-forms and \ \textit{curvature bivectors}. The
curvature 2-forms satisfy Cartan's second structure equation and the curvature
bivectors satisfy equations in analogy with equations of gauge theories. This
immediately suggest to write Einstein's theory in that formalism, something
that has already been done and extensively studied in the past, but not with
the methods used in this paper. However, we did not enter into the details of
that theory in this paper. We only discussed the relation between the
nonhomogeneous $Sl(2,\mathbb{C})$ \ gauge equation satisfied by the curvature
bivector and some other mathematically equivalent formulations of Einstein's
field equations and also we carefully analyzed the relation of nonhomogeneous
$Sl(2,\mathbb{C})$ \ gauge equation satisfied by the curvature bivector and
some interesting equations that appears in M. Sachs \ `unified' field theory
as described recently in \cite{s3} \ and originally introduced in \cite{s1}.
In these books the interested reader may also find a complete list of
references to Sachs papers published in the last 40 years. Next, taking profit
of the mathematical \ methods introduced in this paper, we discussed also some
issues concerning the very important problem of the \ energy-momentum
\ `conservation' in General Relativity, which has defied scientists for almost
all the twenty century, since Hilbert and Einstein found the field equations
of the gravitational field.

Finally, we present in Appendix A the main results of the Clifford bundle
formalism, and in Appendix B we detail the description of Einstein's equations
for the tetrad fields, using the Clifford bundle formalism, where some very
nice operators, which \ have no analogy on classical differential geometry are
exhibit.\medskip

\textbf{Acknowledgment. }Authors are grateful to Dr. Ricardo A. Mosna for very
useful observations, and to a referee for very important comments.

\appendix

\section{\ Clifford Bundles $\mathcal{C\ell}(T^{\ast}M)$ and $\mathcal{C\ell
}(TM)$}

Let $\mathcal{L}=(M,g,D,\tau_{g},\uparrow)$ be a Lorentzian spacetime. This
means that $(M,g,\tau_{g},\uparrow)$ is a four dimensional Lorentzian
manifold, time oriented by $\uparrow$ and spacetime oriented by $\tau_{g}$,
\ and in general $\ M\neq\mathbb{R}^{4}$. Also, $g\in\sec(T^{\ast}M\times
T^{\ast}M)$ is a Lorentzian metric of signature (1,3). $T^{\ast}M$ [$TM$] is
the cotangent [tangent] bundle. $T^{\ast}M=\cup_{x\in M}T_{x}^{\ast}M$,
$TM=\cup_{x\in M}T_{x}M$, and $T_{x}M\simeq T_{x}^{\ast}M\simeq\mathbb{R}%
^{1,3}$, where $\mathbb{R}^{1,3}$ is the Minkowski vector space \cite{sw}. $D$
is the Levi-Civita connection of $g$, i.e\textit{.\/}, $Dg=0$, $\mathcal{R}%
(D)=0$. Also $\Theta(D)=0$, $\mathcal{R}$ and $\Theta$ being respectively the
torsion and curvature tensors. Now, the Clifford bundle of differential forms
$\mathcal{C\ell}(T^{\ast}M)$ is the bundle of algebras\footnote{We can show
using the definitions of section 5 that $\mathcal{C}\ell(T^{\ast}M)$ is a
vector bundle associated with the \emph{\ orthonormal frame bundle}, i.e.,
$\mathcal{C}\!\ell(M)$ $=P_{SO_{+(1,3)}}\times_{ad}Cl_{1,3}$.\ Details about
this construction can be found, e.g., in \cite{moro}.} $\mathcal{C\ell
}(T^{\ast}M)=\cup_{x\in M}\mathcal{C\ell}(T_{x}^{\ast}M)$ , where $\forall
x\in M,\mathcal{C\ell}(T_{x}^{\ast}M)=\mathbb{R}_{1,3}$, the so-called
\emph{spacetime} \emph{algebra }\cite{lounesto}. Locally as a linear space
over the real field $R$, $\mathcal{C\ell}(T_{x}^{\ast}M)$ is isomorphic to the
Cartan algebra $%
{\displaystyle\bigwedge}
(T_{x}^{\ast}M)$ of the cotangent space and $%
{\displaystyle\bigwedge}
T_{x}^{\ast}M=\sum_{k=0}^{4}%
{\displaystyle\bigwedge}
{}^{k}T_{x}^{\ast}M$, where $%
{\displaystyle\bigwedge\nolimits^{k}}
T_{x}^{\ast}M$ is the $\binom{4}{k}$-dimensional space of $k$-forms. The
Cartan bundle $%
{\displaystyle\bigwedge}
T^{\ast}M=\cup_{x\in M}%
{\displaystyle\bigwedge}
T_{x}^{\ast}M$ can then be thought \cite{lami} as \textquotedblleft
imbedded\textquotedblright\ in $\mathcal{C\ell}(T^{\ast}M)$. In this way
sections of $\mathcal{C\ell}(T^{\ast}M)$ can be represented as a sum of
nonhomogeneous differential forms. Let $\{\mathbf{e}_{\mathbf{a}}\}\in\sec
TM,(\mathbf{a}=0,1,2,3)$ be an orthonormal basis $g(\mathbf{e}_{\mathbf{a}%
},\mathbf{e}_{\mathbf{b}})=\eta_{\mathbf{ab}}=\mathrm{diag}(1,-1,-1,-1)$ and
let $\{\theta^{\mathbf{a}}\}\in\sec%
{\displaystyle\bigwedge\nolimits^{1}}
T^{\ast}M\hookrightarrow\sec\mathcal{C\ell}(T^{\ast}M)$ be the dual basis.
Moreover, we denote by $g^{-1}$ the metric in the cotangent bundle.

An analogous construction can be done for the tangent space. The corresponding
Clifford bundle is denoted $\mathcal{C\ell}(TM)$ and their sections are called
multivector fields. All formulas presented below for $\mathcal{C\ell}(T^{\ast
}M)$ have corresponding ones in $\mathcal{C\ell}(TM)$ and this fact has been
used in the text.

\subsection{Clifford product, scalar contraction and exterior products}

The fundamental \emph{Clifford product\footnote{If the reader need more detail
on the Clifford calculus of multivetors he may consult, e.g.,
\cite{femoro101,femoro201,femoro301,femoro401,femoro501,femoro601,femoro701}%
.}} (in what follows to be denoted by juxtaposition of symbols) is generated
by $\theta^{\mathbf{a}}\theta^{\mathbf{b}}+\theta^{\mathbf{b}}\theta
^{\mathbf{a}}=2\eta^{\mathbf{ab}}$ and if $\mathcal{C}\in\sec\mathcal{C\ell
}(T^{\ast}M)$ we have \ref{moro,28}%

\begin{equation}
\mathcal{C}=s+v_{\mathbf{a}}\theta^{\mathbf{a}}+\frac{1}{2!}b_{\mathbf{cd}%
}\theta^{\mathbf{c}}\theta^{\mathbf{d}}+\frac{1}{3!}a_{\mathbf{abc}}%
\theta^{\mathbf{a}}\theta^{\mathbf{b}}\theta^{\mathbf{c}}+p\theta^{\mathbf{5}%
}\;, \label{a.1}%
\end{equation}
where $\theta^{\mathbf{5}}=\theta^{0}\theta^{\mathbf{1}}\theta^{\mathbf{2}%
}\theta^{\mathbf{3}}$ is the volume element and $s$, $v_{\mathbf{a}}$,
$b_{\mathbf{cd}}$, $a_{\mathbf{abc}}$, $p\in\sec%
{\displaystyle\bigwedge\nolimits^{0}}
T^{\ast}M\subset\sec\mathcal{C\ell}(T^{\ast}M)$.

Let $A_{r},\in\sec%
{\displaystyle\bigwedge\nolimits^{r}}
T^{\ast}M\hookrightarrow\sec\mathcal{C\ell}(T^{\ast}M),B_{s}\in\sec%
{\displaystyle\bigwedge\nolimits^{s}}
T^{\ast}M\hookrightarrow\sec\mathcal{C\ell}(T^{\ast}M)$. For $r=s=1$, we
define the \emph{scalar product} as follows:

For $a,b\in\sec%
{\displaystyle\bigwedge\nolimits^{1}}
T^{\ast}M\hookrightarrow\sec\mathcal{C\ell}(T^{\ast}M),$%
\begin{equation}
a\cdot b=\frac{1}{2}(ab+ba)=g^{-1}(a,b). \label{a.2}%
\end{equation}
We also define the \emph{exterior product} ($\forall r,s=0,1,2,3)$ by
\begin{align}
A_{r}\wedge B_{s}  &  =\langle A_{r}B_{s}\rangle_{r+s},\nonumber\\
A_{r}\wedge B_{s}  &  =(-1)^{rs}B_{s}\wedge A_{r} \label{a.3}%
\end{align}
where $\langle\rangle_{k}$ is the component in the subspace $%
{\displaystyle\bigwedge\nolimits^{k}}
T^{\ast}M$ of the Clifford field. The exterior product is extended by
linearity to all sections of $\mathcal{C}\ell(T^{\ast}M).$

For $A_{r}=a_{1}\wedge...\wedge a_{r},B_{r}=b_{1}\wedge...\wedge b_{r}$, the
\textit{scalar product} is defined as
\begin{align}
A_{r}\cdot B_{r}  &  =(a_{1}\wedge...\wedge a_{r})\cdot(b_{1}\wedge...\wedge
b_{r})\nonumber\\
&  =\det\left[
\begin{array}
[c]{ccc}%
a_{1}\cdot b_{1} & \ldots & a_{1}\cdot b_{k}\\
\ldots & \ldots & \ldots\\
a_{k}\cdot b_{1} & \ldots & a_{k}\cdot b_{k}%
\end{array}
\right]  . \label{a.4}%
\end{align}

We agree that if $r=s=0$, the scalar product is simple the ordinary product in
the real field.

Also, if $r\neq s,$ $A_{r}\cdot B_{s}=0$ .

For $r\leq s,A_{r}=a_{1}\wedge...\wedge a_{r},B_{s}=b_{1}\wedge...\wedge
b_{s\text{ }}$we define the \textit{left contraction} by
\begin{equation}
\lrcorner:(A_{r},B_{s})\mapsto A_{r}\lrcorner B_{s}=%
{\displaystyle\sum\limits_{i_{1}<...<i_{r}}}
\epsilon_{1......s}^{i_{1}.....i_{s}}(a_{1}\wedge...\wedge a_{r}%
)\cdot(b_{i_{1}}\wedge...\wedge b_{i_{r}})^{\sim}b_{i_{r}+1}\wedge...\wedge
b_{i_{s}}, \label{a.5}%
\end{equation}
\ where $\sim$ denotes the reverse mapping (\emph{reversion})
\begin{equation}
\sim:\sec%
{\displaystyle\bigwedge\nolimits^{p}}
T^{\ast}M\ni a_{1}\wedge...\wedge a_{p}\mapsto a_{p}\wedge...\wedge a_{1},
\label{a.6}%
\end{equation}
and extended by linearity to all sections of $\mathcal{C\ell}(T^{\ast}M)$. We
agree that for $\alpha,\beta\in\sec%
{\displaystyle\bigwedge\nolimits^{0}}
T^{\ast}M$ the contraction is the ordinary (pointwise) product in the real
field and that if $\alpha\in\sec%
{\displaystyle\bigwedge\nolimits^{0}}
T^{\ast}M$, $A_{r},\in\sec%
{\displaystyle\bigwedge\nolimits^{r}}
T^{\ast}M,B_{s}\in\sec%
{\displaystyle\bigwedge\nolimits^{s}}
T^{\ast}M$ then $(\alpha A_{r})\lrcorner B_{s}=A_{r}\lrcorner(\alpha B_{s})$.
Left contraction is extended by linearity to all pairs of elements of sections
of $\mathcal{C\ell}(T^{\ast}M)$, i.e., for $A,B\in\sec\mathcal{C\ell}(T^{\ast
}M)$%

\begin{equation}
A\lrcorner B=\sum_{r,s}\langle A\rangle_{r}\lrcorner\langle B\rangle
_{s},\text{ }r\leq s. \label{a.7}%
\end{equation}

It is also necessary to introduce in $\mathcal{C\ell}(T^{\ast}M)$ the operator
of \emph{right contraction} denoted by $\llcorner$. The definition is obtained
from the one presenting the left contraction with the imposition that $r\geq
s$ and taking into account that now if $A_{r},\in\sec%
{\displaystyle\bigwedge\nolimits^{r}}
T^{\ast}M,B_{s}\in\sec%
{\displaystyle\bigwedge\nolimits^{s}}
T^{\ast}M$ then $A_{r}\llcorner(\alpha B_{s})=(\alpha A_{r})\llcorner B_{s}$.

\subsection{Some useful formulas}

The main formulas used in the Clifford calculus in the main text can be
obtained from the following ones, where $a\in\sec%
{\displaystyle\bigwedge\nolimits^{1}}
T^{\ast}M$ and $A_{r},\in\sec%
{\displaystyle\bigwedge\nolimits^{r}}
T^{\ast}M,B_{s}\in\sec%
{\displaystyle\bigwedge\nolimits^{s}}
T^{\ast}M$:
\begin{align}
aB_{s}  &  =a\lrcorner B_{s}+a\wedge B_{s},B_{s}a=B_{s}\llcorner a+B_{s}\wedge
a,\label{a.8}\\
a\lrcorner B_{s}  &  =\frac{1}{2}(aB_{s}-(-)^{s}B_{s}a),\nonumber\\
A_{r}\lrcorner B_{s}  &  =(-)^{r(s-1)}B_{s}\llcorner A_{r},\nonumber\\
a\wedge B_{s}  &  =\frac{1}{2}(aB_{s}+(-)^{s}B_{s}a),\nonumber\\
A_{r}B_{s}  &  =\langle A_{r}B_{s}\rangle_{|r-s|}+\langle A_{r}\lrcorner
B_{s}\rangle_{|r-s-2|}+...+\langle A_{r}B_{s}\rangle_{|r+s|}\nonumber\\
&  =\sum\limits_{k=0}^{m}\langle A_{r}B_{s}\rangle_{|r-s|+2k},\text{ }%
m=\frac{1}{2}(r+s-|r-s|),\nonumber\\
A_{r}\lrcorner B_{r}  &  =A_{r}\llcorner B_{r}=\tilde{A}_{r}\cdot B_{r}%
=A_{r}\cdot\tilde{B}_{r}\nonumber
\end{align}

\subsection{Hodge star operator}

Let $\star$ be the usual Hodge star operator $\star:%
{\displaystyle\bigwedge\nolimits^{k}}
T^{\ast}M\rightarrow%
{\displaystyle\bigwedge\nolimits^{4-k}}
T^{\ast}M$. If $B\in\sec%
{\displaystyle\bigwedge\nolimits^{k}}
T^{\ast}M$, $A\in\sec%
{\displaystyle\bigwedge\nolimits^{4-k}}
T^{\ast}M$ and $\tau\in\sec%
{\displaystyle\bigwedge\nolimits^{4}}
T^{\ast}M$ is the volume form, then $\star B$ is defined by
\[
A\wedge\star B=(A\cdot B)\tau.
\]

Then we can show that if $A_{p}\in\sec%
{\displaystyle\bigwedge\nolimits^{p}}
T^{\ast}M\hookrightarrow\sec\mathcal{C\!\ell}(T\ast M)$ we have
\begin{equation}
\star A_{p}=\widetilde{A_{p}}\theta^{\mathbf{5}}. \label{a.hodge}%
\end{equation}
This equation is enough to prove very easily the following identities (which
are used in the main text):%
\begin{align}
A_{r}\wedge\star B_{s}  &  =B_{s}\wedge\star A_{r};\hspace{0.15in}%
r=s,\nonumber\\
A_{r}\lrcorner\star B_{s}  &  =B_{s}\lrcorner\star A_{r};\hspace
{0.15in}r+s=4,\nonumber\\
A_{r}\wedge\star B_{s}  &  =(-1)^{r(s-1)}\star(\tilde{A}_{r}\lrcorner
B_{s});\hspace{0.15in}r\leq s,\nonumber\\
A_{r}\lrcorner\star B_{s}  &  =(-1)^{rs}\star(\tilde{A}_{r}\wedge
B_{s});\hspace{0.15in}r+s\leq4 \label{Aidentities}%
\end{align}

Let $d$ and $\delta$ be respectively the differential and Hodge codifferential
operators acting on sections of $%
{\displaystyle\bigwedge}
T^{\ast}M$. If $\omega_{p}\in\sec%
{\displaystyle\bigwedge\nolimits^{p}}
T^{\ast}M\hookrightarrow\sec\mathcal{C\ell}(T^{\ast}M)$, then $\delta
\omega_{p}=(-)^{p}\star^{-1}d\star\omega_{p}$, where $\star^{-1}%
\star=\mathrm{identity}$. When applied to a $p$-form we have%
\[
\star^{-1}=(-1)^{p(4-p)+1}\star\hspace{0.15in}.
\]

\subsection{Action of $D_{\mathbf{e}_{\mathbf{a}}}$ on Sections of
$\mathcal{C\!\ell}(TM)$ and $\mathcal{C\!\ell}(T^{\ast}M)$}

Let $D_{\mathbf{e}_{\mathbf{a}}}$ be a metrical compatible covariant
derivative operator acting on sections of the tensor bundle. It can be easily
shown (see, e.g., \cite{cru}) that $D_{\mathbf{e}_{\mathbf{a}}}$ is also a
covariant derivative operator on the Clifford bundles $\mathcal{C\!\ell}(TM)$
and $\mathcal{C\!\ell}(T^{\ast}M).$

Now, if \ $A_{p}\in\sec%
{\displaystyle\bigwedge\nolimits^{p}}
T^{\ast}M\hookrightarrow\sec\mathcal{C\!\ell}(M)$ \ we can show, very easily
by explicitly performing the calculations\footnote{A derivation of this
formula from the genral theory of connections can be found in \cite{moro}.}
that%
\begin{equation}
D_{\mathbf{e}_{\mathbf{a}}}A_{p}=\partial_{\mathbf{e}_{\mathbf{a}}}A_{p}%
+\frac{1}{2}[\omega_{\mathbf{e}_{\mathbf{a}}},A_{p}], \label{der1}%
\end{equation}
where the $\omega_{\mathbf{e}_{\mathbf{a}}}\in\sec%
{\displaystyle\bigwedge\nolimits^{2}}
T^{\ast}M\hookrightarrow\sec\mathcal{C\!\ell}(M)$ may be called
\textit{Clifford} \textit{connection 2-forms. }They\textit{ }are given
by:\textit{ }%
\begin{equation}
\omega_{\mathbf{e}_{\mathbf{a}}}=\frac{1}{2}\omega_{\mathbf{a}}^{\mathbf{bc}%
}\theta_{\mathbf{b}}\theta_{\mathbf{c}}=\frac{1}{2}\omega_{\mathbf{a}%
}^{\mathbf{bc}}\theta_{\mathbf{b}}\wedge\theta_{\mathbf{c}}, \label{der2}%
\end{equation}
where (in standard notation)%
\begin{equation}
D_{\mathbf{e}_{\mathbf{a}}}\theta_{\mathbf{b}}=\omega_{\mathbf{ab}%
}^{\mathbf{c}}\theta_{\mathbf{c}},\hspace{0.15in}D_{\mathbf{e}_{\mathbf{a}}%
}\theta^{\mathbf{b}}=-\omega_{\mathbf{ac}}^{\mathbf{b}}\theta^{\mathbf{c}%
},\hspace{0.15cm}\omega_{\mathbf{a}}^{\mathbf{bc}}=-\omega_{\mathbf{a}%
}^{\mathbf{cb}} \label{der3}%
\end{equation}

An analogous formula to Eq.(\ref{der1}) is valid for the covariant derivative
of sections of $\mathcal{C\!\ell}(TM)$ and they were used in several places in
the main text.

\subsection{Dirac Operator, Differential and Codifferential}

The Dirac operator acting on sections of $\mathcal{C\!\ell}(T^{\ast}M)$ is the
invariant first order differential operator
\begin{equation}
{%
\mbox{\boldmath$\partial$}%
}=\theta^{\mathbf{a}}D_{\mathbf{e}_{\mathbf{a}}}, \label{1.5}%
\end{equation}
and we can show (see, e.g., \cite{rq}) that when $D_{\mathbf{e}_{\mathbf{a}}}$
is the Levi-Civita covariant derivative operator, the following important
result holds:
\begin{equation}
{%
\mbox{\boldmath$\partial$}%
}={%
\mbox{\boldmath$\partial$}%
}\wedge\,+\,{%
\mbox{\boldmath$\partial$}%
}\lrcorner=d-\delta. \label{1.6}%
\end{equation}

The square of the Dirac operator ${%
\mbox{\boldmath$\partial$}%
}^{2}$ is called the \textit{Hodge Laplacian}. We have
\begin{equation}
{%
\mbox{\boldmath$\partial$}%
}^{2}=-(d\delta+\delta d). \label{1.6bis}%
\end{equation}
This operator is not to be confused with the \textit{covariant D'Alembertian}
which is given by%
\begin{equation}
\square={%
\mbox{\boldmath$\partial$}%
\cdot%
\mbox{\boldmath$\partial$}%
.} \label{1.6a}%
\end{equation}
The following identities were used in the text%
\begin{align}
dd  &  =\delta\delta=0,\nonumber\\
d{%
\mbox{\boldmath$\partial$}%
}^{2}  &  ={%
\mbox{\boldmath$\partial$}%
}^{2}d;\hspace{0.15in}\delta{%
\mbox{\boldmath$\partial$}%
}^{2}={%
\mbox{\boldmath$\partial$}%
}^{2}\delta,\nonumber\\
\delta\star &  =(-1)^{p+1}\star d;\hspace{0.15in}\star\delta=(-1)^{p}\star
d,\nonumber\\
d\delta\star &  =\star d\delta;\hspace{0.15in}\star d\delta=\delta
d\star;\hspace{0.15in}\star{%
\mbox{\boldmath$\partial$}%
}^{2}={%
\mbox{\boldmath$\partial$}%
}^{2}\star\label{A.identities1}%
\end{align}

\subsection{Maxwell Equation}

Maxwell equations in the Clifford bundle of differential forms resume in one
single equation. Indeed, if $F\in\sec%
{\displaystyle\bigwedge\nolimits^{2}}
T^{\ast}M\hookrightarrow\sec\mathcal{C\ell}(T^{\ast}M)$ is the electromagnetic
field and $J_{e}\in\sec%
{\displaystyle\bigwedge\nolimits^{1}}
T^{\ast}M\hookrightarrow\sec\mathcal{C\ell}(T^{\ast}M)$ is the electromagnetic
current, we have Maxwell equation\footnote{Then, there is no misprint in the
title of this section.}:
\begin{equation}
{%
\mbox{\boldmath$\partial$}%
}F=J_{e}. \label{1.7}%
\end{equation}

Eq.(\ref{1.7}) is equivalent to the pair of equations%
\begin{align}
dF  &  =0,\label{1.8a}\\
\delta F  &  =-J_{e}. \label{1.8b}%
\end{align}

Eq.(\ref{1.8a}) is called the homogenous equation and Eq.(\ref{1.8b}) is
called the nonhomogeneous equation. Note that it can be written also as:%
\begin{equation}
d\star F=-\star J_{e}. \label{1.9}%
\end{equation}

\section{Einstein Field Equations for the Tetrad Fields $\theta^{\mathbf{a}}$}

In the main text we gave a Clifford bundle formulation of the field equations
of general relativity in a form that resembles a $Sl(2,\mathbb{C)}$ gauge
theory and also a formulation in terms of a set of $2$-form fields
$\star\mathcal{S}^{\mathbf{a}}$. Here we want to \textit{recall }yet another
face\textit{ }of Einstein's equations, i.e., we show how to write the field
equations directly for the tetrad fields $\theta^{\mathbf{a}}$ in such a way
that the obtained equations are equivalent to Einstein's field equations. This
is done in order to compare the correct equations for that objects which some
other equations proposed for these objects that appeared recently in the
literature (and which will be discussed below). Before proceeding, we mention
that, of course, we could write analogous (and equivalent) equations for the
dual tetrads $\mathbf{e}_{\mathbf{a}}$.

As shown in details in papers \ \cite{rq,qr} \ the correct wave like equations
satisfied by the $\theta^{\mathbf{a}}$ are\footnote{Of course, there are
analogous equations for the $\mathbf{e}_{\mathbf{a}}$ \cite{hestenes}, where
in that case, the Dirac operator must be defined (in an obvious way) as acting
on sections of the Clifford bundle of multivectors, that has been introduced
in section 3.}:
\begin{equation}
-({%
\mbox{\boldmath$\partial$}%
}\cdot{%
\mbox{\boldmath$\partial$}%
})\theta^{\mathbf{a}}+{%
\mbox{\boldmath$\partial$}%
}\wedge({%
\mbox{\boldmath$\partial$}%
}\cdot\theta^{\mathbf{a}})+{%
\mbox{\boldmath$\partial$}%
}\lrcorner({%
\mbox{\boldmath$\partial$}%
}\wedge\theta^{\mathbf{a}})=\mathcal{T}^{\mathbf{a}}-\frac{1}{2}%
T\theta^{\mathbf{a}}. \label{11.1}%
\end{equation}

In Eq.(\ref{11.1}), $\mathcal{T}^{\mathbf{a}}=T_{\mathbf{b}}^{\mathbf{a}%
}\theta^{\mathbf{b}}\in\sec\bigwedge\nolimits^{1}T^{\ast}M\hookrightarrow
\sec\mathcal{C}\ell(T^{\ast}M)$ are the energy momentum $1$-form fields and
$T=T_{\mathbf{a}}^{\mathbf{a}}=-R=-R_{\mathbf{a}}^{\mathbf{a}}$, with
$T_{\mathbf{ab}}$ the energy momentum tensor of matter. When $\theta
^{\mathbf{a}}$ is an exact differential, and in this case we write
$\theta^{\mathbf{a}}\mapsto$ $\theta^{\mu}=dx^{\mu}$ and if the coordinate
functions are harmonic, i.e., $\delta\theta^{\mu}=-{%
\mbox{\boldmath$\partial$}%
}\theta^{\mu}=0$, Eq.(\ref{11.1}) becomes%
\begin{equation}
\square\theta^{\mu}+\frac{1}{2}R\theta^{\mu}=-\mathcal{T}^{\mu},
\label{11.1BIS}%
\end{equation}
where we have used Eq.(\ref{1.6a}).

In Eq.(\ref{11.1}) ${%
\mbox{\boldmath$\partial$}%
}=\theta^{\mathbf{a}}D_{\mathbf{e}_{a}}={%
\mbox{\boldmath$\partial$}%
}\wedge+$ ${%
\mbox{\boldmath$\partial$}%
}\lrcorner$ $=d-\delta$ is the Dirac (like) operator acting on sections of the
Clifford bundle $\mathcal{C}\ell(T^{\ast}M)$ defined in the previous Appendix.

With these formulas we can write%
\begin{align}
{%
\mbox{\boldmath$\partial$}%
}^{2}  &  ={%
\mbox{\boldmath$\partial$}%
}\cdot{%
\mbox{\boldmath$\partial$}%
}+{%
\mbox{\boldmath$\partial$}%
}\wedge{%
\mbox{\boldmath$\partial$}%
},\nonumber\\
{%
\mbox{\boldmath$\partial$}%
}\wedge{%
\mbox{\boldmath$\partial$}%
}  &  =-{%
\mbox{\boldmath$\partial$}%
}\cdot{%
\mbox{\boldmath$\partial$}%
}+{%
\mbox{\boldmath$\partial$}%
}\wedge{%
\mbox{\boldmath$\partial$}%
}\lrcorner+{%
\mbox{\boldmath$\partial$}%
}\lrcorner{%
\mbox{\boldmath$\partial$}%
}\wedge,\hspace{0.15in} \label{11.2}%
\end{align}
with%
\begin{align}
{%
\mbox{\boldmath$\partial$}%
}\cdot{%
\mbox{\boldmath$\partial$}%
}  &  =\eta^{\mathbf{ab}}(D_{\mathbf{e}_{\mathbf{a}}}D_{\mathbf{e}%
_{\mathbf{b}}}-\omega_{\mathbf{ab}}^{\mathbf{c}}D_{\mathbf{e}_{\mathbf{c}}%
}),\nonumber\\
{%
\mbox{\boldmath$\partial$}%
}\wedge{%
\mbox{\boldmath$\partial$}%
}  &  =\theta^{\mathbf{a}}\wedge\theta^{\mathbf{b}}(D_{\mathbf{e}_{\mathbf{a}%
}}D_{\mathbf{e}_{\mathbf{b}}}-\omega_{\mathbf{ab}}^{\mathbf{c}}D_{\mathbf{e}%
_{\mathbf{c}}}). \label{11.3}%
\end{align}

Note that $D_{e_{a}}\theta^{\mathbf{b}}=-\omega_{\mathbf{ac}}^{\mathbf{b}%
}\theta^{\mathbf{c}}$ and a \ somewhat long, but simple calculation
\footnote{The calculation is done in detail in \cite{rq,qr}.} shows that%
\begin{equation}
({%
\mbox{\boldmath$\partial$}%
}\wedge{%
\mbox{\boldmath$\partial$}%
})\theta^{\mathbf{a}}=\mathcal{R}^{\mathbf{a}},\label{11.5}%
\end{equation}
where, as already defined, $\mathcal{R}^{\mathbf{a}}=R_{\mathbf{b}%
}^{\mathbf{a}}\theta^{\mathbf{b}}$ are the Ricci 1-forms. We also observe
(that for the best of our knowledge) ${%
\mbox{\boldmath$\partial$}%
}\wedge{%
\mbox{\boldmath$\partial$}%
}$ that has been named the Ricci operator in \cite{qr} has no analogue \ in
classical differential geometry.

Note that Eq.(\ref{11.1}) can be written after some algebra as
\begin{equation}
\mathcal{R}^{\mu}-\frac{1}{2}R\theta^{\mu}=\mathcal{T}^{\mu}, \label{11.6}%
\end{equation}
with $\mathcal{R}^{\mu}=R_{\nu}^{\mu}dx^{\nu}$ and $\mathcal{T}^{\mu}=T_{\nu
}^{\mu}dx^{\nu}$, $\theta^{\mu}=dx^{\mu}$ in a coordinate chart of the maximal
atlas of $M$ covering an open set $U\subset M$.

We are now prepared to make some crucial comments concerning some recent
papers (\cite{cleevans},\cite{e1}-\cite{e4}).

(\textbf{i}) In (\cite{cleevans},\cite{e1}-\cite{e4}) authors claims that the
$\mathbf{e}_{\mathbf{a}}$, $\mathbf{a}=0,1,2,3$ satisfy the equations\medskip

\begin{center}%
\begin{tabular}
[c]{|l|}\hline
$(\square+T)\mathbf{e}_{\mathbf{a}}=0.$\\\hline
\end{tabular}
\medskip\ 
\end{center}

They thought to have produced a valid derivation for that equations. We will
not comment on that derivation here. Enough is to say that if that equation
was true it would imply that $(\square+T)\theta^{\mathbf{a}}=0$. This is not
the case. Indeed, as a careful reader may verify, the true equation satisfied
by any one of the $\theta^{\mathbf{a}}$ is Eq.(\ref{11.1}).

(\textbf{ii}) We quote that author of \cite{e1,e2,e3} explicitly wrote several
times that the "electromagnetic potential"\footnote{In \cite{e1,e2,e3} author
do identify their "electromagnetic potential" with the \ bivector valued
connection 1-form ${%
\mbox{\boldmath$\omega$}%
}$ that we introduced in section above. As we explained with details this
cannot be done because that quantity is related to gravitation, not
electromagnetism.} $\mathbf{A}$ in their theory (a 1-form with values in a
vector space) satisfies the following wave equation,\medskip

\begin{center}%
\begin{tabular}
[c]{|c|}\hline
$(\square+T)\mathbf{A}=0.$\\\hline
\end{tabular}
\medskip
\end{center}

Now, this equation cannot be correct even for the usual $U(1)$ gauge potential
of classical electrodynamics \footnote{Which must be one of the gauge
components of the gauge field.} $A\in\sec%
{\displaystyle\bigwedge\nolimits^{1}}
T^{\ast}M\subset\sec\mathcal{C\ell(}T^{\ast}M)$. Indeed, in vacuum Maxwell
equation reads (see Eq.(\ref{1.7}))%
\begin{equation}
{%
\mbox{\boldmath$\partial$}%
}F=0, \label{11.9}%
\end{equation}
where $F={%
\mbox{\boldmath$\partial$}%
}A={%
\mbox{\boldmath$\partial$}%
}\wedge A=dA$, \textit{if} we work in the Lorenz gauge ${%
\mbox{\boldmath$\partial$}%
}\cdot A={%
\mbox{\boldmath$\partial$}%
}\lrcorner A=-\delta A=0$. \ Now, since we have according to Eq.(\ref{1.6bis})
that ${%
\mbox{\boldmath$\partial$}%
}^{2}=-(d\delta+\delta d),$we get
\begin{equation}
{%
\mbox{\boldmath$\partial$}%
}^{2}A=0. \label{11.11}%
\end{equation}

A simple calculation then shows that in the coordinate basis introduced above
we have,
\begin{equation}
({%
\mbox{\boldmath$\partial$}%
}^{2}A)_{\alpha}=g^{\mu\nu}D_{\mu}D_{\nu}A_{\alpha}+R_{\alpha}^{\nu}A_{\nu}
\label{11.12}%
\end{equation}
and we see that Eq.(\ref{11.11}) reads in components\footnote{Take into
account that in Einstein theory the term $R_{\mu}^{\nu}A_{\nu}=0$ in vacuum.}%
\begin{equation}
\square A_{\mu}+R_{\mu}^{\nu}A_{\nu}=0. \label{11.13}%
\end{equation}

Eq.(\ref{11.13}) can be found, e.g., in Eddington's book \ \cite{eddington} on
page 175.

Finally we make a single comment on reference \cite{cleevans}, because this
paper is related to Sachs\ `unified' theory in the sense that authors try to
identify Sachs `electromagnetic' field (discussed in the main text) with a
supposedly existing longitudinal electromagnetic field predict by their
theory. Well, on \cite{cleevans} we can read at the beginning of section 1.1:

\textquotedblleft The antisymmetrized form of special relativity [1] has
spacetime metric given by the enlarged structure%
\begin{equation}
\eta^{\mu\nu}=\frac{1}{2}\left(  \sigma^{\mu}\sigma^{\nu\ast}+\sigma^{\nu
}\sigma^{\mu\ast}\right)  , \tag{1.1.}%
\end{equation}
where $\sigma^{\mu}$ are the Pauli matrices satisfying a clifford (sic)
algebra
\[
\{\sigma^{\mu},\sigma^{\nu}\}=2\delta^{\mu\nu},
\]
which are represented by
\begin{equation}
\sigma^{0}=\left(
\begin{array}
[c]{cc}%
1 & 0\\
0 & 1
\end{array}
\right)  ,\sigma^{1}=\left(
\begin{array}
[c]{cc}%
0 & 1\\
1 & 0
\end{array}
\right)  ,\sigma^{2}=\left(
\begin{array}
[c]{cc}%
0 & -i\\
i & 0
\end{array}
\right)  ,\sigma^{3}=\left(
\begin{array}
[c]{cc}%
1 & 0\\
0 & -1
\end{array}
\right)  . \tag{1.2}%
\end{equation}
The $\ast$ operator denotes quaternion conjugation, which translates to a
spatial parity transformation.\textquotedblright

Well, we comment as follows: the $\ast$ is not really defined anywhere in
\cite{cleevans}. If it refers to a spatial parity operation, we infer that
$\sigma^{0\ast}=\sigma^{0}$ and \ $\sigma^{i\ast}=-\sigma^{i}$. Also,
$\eta^{\mu\nu}$ is not defined, but Eq.(3.5) of \cite{cleevans} make us to
infers that $\eta^{\mu\nu}=$diag$(1,-1,-1,1)$. In that case Eq.(1.1) above
(with the first member understood as $\eta^{\mu\nu}\sigma^{0}$) is true but
the equation $\{\sigma^{\mu},\sigma^{\nu}\}=2\delta^{\mu\nu}$ is false. Enough
is to see that \ $\{\sigma^{0},\sigma^{i}\}=2\sigma^{i}$ $\neq2\delta^{0i}$. \

\end{document}